\documentclass[final,5p,times,twocolumn,numbers,sort&compress]{elsarticle}

\usepackage{amssymb}
\usepackage{lipsum}
\usepackage{float}
\usepackage{units}
\usepackage{slashed}
\usepackage{multirow}
\usepackage{amsmath}
\usepackage{graphicx}
\usepackage{subcaption}
\usepackage{tikz}
\usepackage[compat=1.1.0]{tikz-feynman}
\tikzfeynmanset{warn luatex=false}

\usepackage{tabularx}
\usepackage{booktabs}  

\usepackage{xcolor}
\usepackage[normalem]{ulem}

\journal{International Journal of Modern Physics A}

\begin{document}

\begin{frontmatter}



\title{\large Unveiling stealth SUSY in the low missing energy terrain at the LHC}


\author[inst1,inst2]{Justin I. Alvarez\corref{cor1}}
\ead{jalvarez@nip.upd.edu.ph}
\author[inst1]{Marvin M. Flores}
\cortext[cor1]{Corresponding author}
\address[inst1]{National Institute of Physics, University of the Philippines - Diliman, Quezon City 1104, Philippines}
\address[inst2]{Philippine Science High School - Main Campus, Quezon City 1101, Philippines}

\begin{abstract}
Using a set of missing energy-independent selection criteria involving large-radius jet multiplicity and angular separation among the final states, we extend the discovery potential of the  stealth SUSY squark-pair production in the low missing transverse energy $(\slashed{E}_T)$ parameter space, thereby complementing existing ATLAS and CMS searches that are sensitive in the high $\slashed{E}_T$ regime. With these kinematic cuts, the estimated sensitivity was extended to squark masses up to $m_{\tilde{q}} = \unit[2500]{GeV}$ for a bino mass of $m_{\tilde{\chi}^0_1} = \unit[650]{GeV}$.
\end{abstract}



\begin{keyword}
Supersymmetry \sep Phenomenology



\end{keyword}

\end{frontmatter}




\section{Introduction}
Among the existing beyond-the-Standard-Model (BSM) scenarios, supersymmetry (SUSY) stands out as a leading theoretical framework, offering solutions to phenomena that the Standard Model cannot explain, such as the hierarchy problem, unification of the fundamental forces and the nature of dark matter \cite{drees2005theory}. Hence, the search for SUSY is one of the goals of the Large Hadron Collider (LHC) experiment located at CERN, Geneva, Switzerland. Currently, the LHC experiment is in its Run 3 and no evidence of SUSY particles has been observed as of writing. 

The typical signature for SUSY involves a large missing transverse energy $(\slashed{E}_T)$. This approach is motivated by R-parity conservation, which implies that the lightest supersymmetric particle (LSP) is stable and contributes to the missing transverse energy \cite{fan2011stealth, chatrchyan2013search}. The LSP, often a neutralino, is a possible dark matter candidate because it is neutral and weakly interacting, which prevents it from leaving signatures in the detector \cite{chatrchyan2013search, cms2014search}. In light of recent results after Run 2 of the LHC experiment, higher bounds on squark and gluino masses have been set for SUSY searches involving high $\slashed{E}_T$ signature \cite{cms2021_dilepton_susy, cms2020_boostedZ_susy, cms2020_ss_multileptons}. To avoid abandoning the SUSY framework, some of its proponents introduced a modified model called stealth Supersymmetry (stealth SUSY), where the minimal supersymmetric standard model (MSSM) is augmented with a hidden sector \cite{fan2011stealth, fan2016stealth}. 

\begin{figure}[htbp]
  \centering
  \includegraphics[width = 0.7\linewidth]{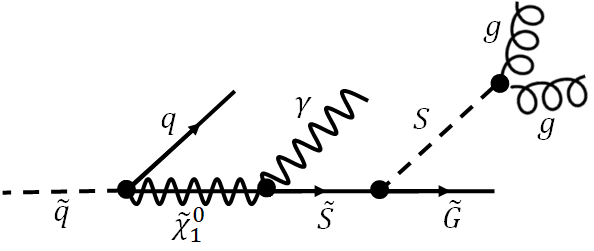}
  \caption{Squark decay in the stealth SUSY model}
  \label{fig:cms2013}
\end{figure}

The simplest stealth SUSY model assumes low-scale SUSY breaking that introduces a hidden sector of particles at the weak scale, which experiences minimal SUSY breaking through the interactions with SM fields \cite{cms2014search}. A sample decay chain is shown in Figure~\ref{fig:cms2013}. In the stealth SUSY  model, the standard LSP $(\tilde{\chi}^0_1)$ takes on a new role as the lighest ``visible sector" particle (LVSP). In the literature this is often referred to as ``bino'', a convention that will also be adopted in this study \cite{fan2011stealth}.

\begin{figure}[H]
  \centering
  \begin{minipage}{0.35\linewidth}
    \centering
    \includegraphics[width=\linewidth]{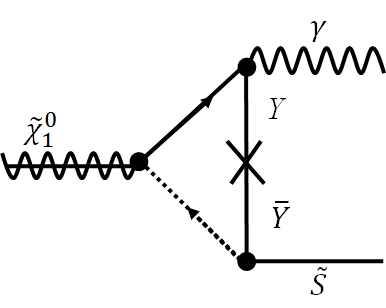}
    \caption*{(a)}
  \end{minipage}
  \hspace{0.02\linewidth} 
  \begin{minipage}{0.35\linewidth}
    \centering
    \includegraphics[width=\linewidth]{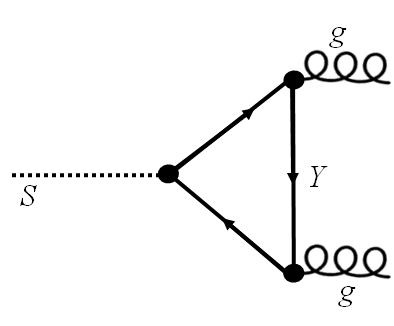}
    \caption*{(b)}
  \end{minipage}
  \caption{One loop-induced couplings with vector-like states allow the decay of bino to a singlino and photon, as well as the decay of singlet scalar to gluons.}
  \label{fig:loops}
\end{figure}

In the so-called $SY\bar{Y}$ model of stealth SUSY, an appropriate portal for the LVSP to transition into the stealth sector is provided by vector-like states $Y$ and $\bar{Y}$, which are charged under the SM gauge group. They couple to the singlet chiral superfield $S$ through the superpotential \cite{fan2011stealth}: 
\begin{equation}
W = \lambda SY \bar{Y} + \frac{m_S}{2} S^2 + m_Y Y \bar{Y}    
\end{equation}

\noindent
where $\lambda$ denotes the coupling between the singlet $S$ and the vector-like messenger fields $Y$ and $\bar{Y}$. The parameters $m_S$ and $m_Y$ represent the supersymmetric mass terms for $S$ and the vector-like pairs $Y$, and $\bar{Y}$, respectively. Since $Y$ and $\bar{Y}$ are charged under SM, they mediate the interactions between the stealth and visible sectors. In particular, they generate at one loop an effective bino–photon–$\tilde{S}$ vertex, which enables the bino to decay into $\tilde{S}$ while emitting a photon. They also induce effective couplings that allow the scalar $S$ to decay into gluons as shown in Figure~\ref{fig:loops} \cite{flores2020constraining}.

The LVSP can decay into a lighter hidden sector without violating R-parity conservation. In this framework, the LVSP decays into a hidden-sector SUSY particle, the fermionic singlino ($\tilde{S}$) which is the superpartner of the scalar singlet ($S$). Since $S$ is even for R-parity, it can subsequently decay into jets through the process $S \to gg$.

The defining feature of stealth SUSY is that the singlet chiral supermultiplet $\mathbf{S} = (S, \tilde{S})$ is nearly supersymmetric, with only a small mass splitting between its components. Specifically, the mass difference $\delta m = m_{\tilde{S}} - m_{S}$ is much smaller than the singlino mass, $\delta m \ll m_{\tilde{S}}$. The mass degeneracy of the hidden-sector SUSY particle and its partner will result in an LSP with little $\slashed{E}_T$ in its final state, consequently evading standard SUSY searches that depend on large amounts of $\slashed{E}_T$ in the final state \cite{fan2011stealth}. In the case of the decay chain in Figure~\ref{fig:cms2013}, the gravitino ($\tilde{G}$) plays the role of the LSP which carries the little missing transverse energy.

The paper is organized as follows: Section~\ref{sec:RRL} reviews previous studies and searches related to stealth SUSY. Section~\ref{sec:event_gen}
 describes the generation of events for the stealth SUSY signal and background. Section~\ref{sec:prev_searches} compares the latest CMS stealth SUSY exclusion results with earlier searches at $\sqrt{s} = \unit[13]{TeV}$. Section~\ref{sec:analysis} details the numerical analysis, while Section~\ref{sec:signal_region} discusses the Signal Regions used to enhance the signal sensitivity over the dominant background. Finally, Section~\ref{sec:conclusion} summarizes the conclusions.

\section{Previous Studies and Searches on stealth SUSY}
\label{sec:RRL}

In 2011, Fan, Reece, and Ruderman introduced a broad class of supersymmetric models that preserve R-parity but lack missing energy signatures \cite{fan2011stealth}. In their initial work, they highlighted possible discovery modes like highly displaced vertices, triple resonances like $\gamma j j$, and the presence of a very large number of b-jets. They also pointed out that for the three-body decay $\tilde{S} \to S(\rightarrow gg) \tilde{G}$, its decay width $(\Gamma_{\tilde{S}})$ is given by:
\begin{equation}
\Gamma_{\tilde{S}} = \frac{m^5_{\tilde{S}}}{16\pi F^2}\left(1 - \frac{m^2_S}{m^2_{\tilde{S}}} \right) \approx \frac{m^2_{\tilde{S}}(\delta m)^4}{\pi F^2}
\end{equation}

For a SUSY breaking scale, $\sqrt{F} =\unit[100]{TeV}$, $m_{\tilde{S}} = \unit[100]{GeV}$, and $m_{S} = \unit[90]{GeV}$, the resulting decay length is $\unit[8]{cm}$. This means that the singlino $\tilde{S}$, once produced at the collision point, will travel this distance before decaying, well within the tracking volume of ATLAS and CMS detectors. A more detailed study followed in 2012 and 2016 where they discussed other models that realize the concept of stealth SUSY \cite{fan2012_stealth_susy_sampler, fan2016stealth}. They mentioned stealth models that couple through a baryon portal or $Z'$ gauge interactions, and other simplified models of stealth SUSY that motivate $\unit[13]{TeV}$ LHC searches.

The CMS Collaboration has conducted numerous searches that focused on the stealth SUSY model. The first of these analyses is based on $\unit[4.96]{fb^{-1}}$ of proton-proton collision data at $\sqrt{s} =  \unit[7]{TeV}$, collected in 2011 at the LHC \cite{chatrchyan2013search}. They considered the same decay shown in Figure~\ref{fig:cms2013} that includes a degenerate light squark $(\tilde{q})$, a ``bino-like" LVSP $\tilde{\chi}^0_1$, and a hidden sector containing a singlet state $S$ and its fermionic ``singlino" superpartner $\tilde{S}$. The gluino $(\tilde{g})$ is fixed at $\unit[1500]{GeV}$. This study considers the same squark–antisquark $(\tilde{q},\bar{\tilde{q}})$ production process in an earlier work, where squarks decay via $\tilde{q} \to q \tilde{\chi}^0_1$ \cite{lhc2011_simplified_models}. The main differences are the presence of a hidden sector, a characteristic of stealth SUSY, and the participation of gluinos $(\tilde{g})$ in the production mechanism. The paper looks for new phenomena in events with two photons and four or more hadronic jets, with no requirement on missing transverse energy $(\slashed{E}_T)$. They were then able to compute the limit on the stealth SUSY cross section as a function of $M_{\tilde{q}}$. Lastly, squark masses below $\unit[1430]{GeV}$ were excluded by this analysis at 95 $\%$ Confidence Level (CL). This work placed the first experimental limit on the parameters of the stealth SUSY model. 

\begin{figure}[htbp]
  \centering
  \includegraphics[width = 0.7\linewidth]{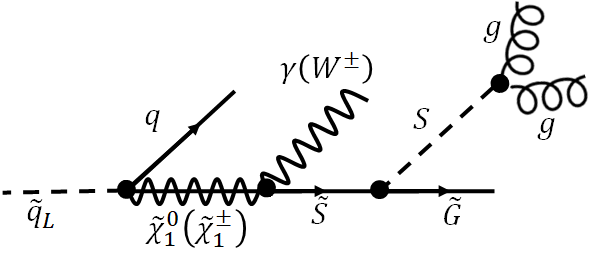}
  \caption{Squark decay to a quark and gaugino in the stealth SUSY model. The gauginos then decay into a singlino $S$ and a $\gamma$ or $W^\pm$ boson}
  \label{fig:cms2015}
\end{figure}

The CMS Collaboration extended this work based on $\unit[19.7]{fb^{-1}}$ integrated luminosity of proton-proton collisions at $\sqrt{s} =  \unit[8]{TeV}$ \cite{cms2014search}. The analysis used data collected with the CMS detector at the LHC in 2012. Figure~\ref{fig:cms2015} shows the decay channel used in this study. Since this study also considered the decay of squarks via $\tilde{\chi}^\pm_1 \to \tilde{S} W^\pm$ for the $l^{\pm}$ analysis, they restricted the analysis on s-channel production of left-handed squarks ($\tilde{u}$, $\tilde{d}$, $\tilde{s}$ and $\tilde{c}$) for consistency. As with the previous study, the $\gamma$ analysis search for  new phenomena in events with four or more jets, low missing transverse momentum, and two photons while the $l^{\pm}$ analysis search for new phenomena on events with four or more jets, low missing transverse momentum, and one electron and one muon of opposite charge via leptonic decay of $W^{\pm}$ boson in the final state. 

At the end of this work, the CMS collaboration was able to set lower limits on squark masses for the $\gamma$ analysis between 700 and 1050 GeV, depending on the neutralino mass at 95 $\%$ CL. For the $l^{\pm}$ analysis, they were able to exclude the masses of squarks below $\unit[550]{GeV}$ at 95 $\%$ CL. This work supersedes the limit of the previous $\gamma$ analysis and also reports the first exclusion limits for the squark masses via the $l^{\pm}$ channel.

On the phenomenology side, a paper published in 2020 considered the same stealth SUSY squark decay chain shown in Figure~\ref{fig:cms2013} \cite{flores2020constraining}. In their work, they reconstructed the bino mass $(m_{\tilde{\chi}^0_1} \approx M(\gamma g g))$ using the kinematic variables available in the detector. To be specific, they presented the following selection criteria:
\begin{enumerate}
    \item The number large-radius jets $(R = 1.0)$ in an event must be greater than 3
    \item The number of high-transverse-momentum photons $(p_T > \unit[200]{GeV})$ in an event must be greater than 1
    \item The photon must be contained inside the large-radius jet $(\Delta R_{j_1,\gamma} < 1)$.
\end{enumerate}

Using these kinematic cuts, this work showed that the sensitivity calculated for two benchmark points in the squark-bino mass phase space has a stronger exclusion potential than the strongest search in ATLAS at $\unit[36]{fb^{-1}}$ in excluding possible squark and bino masses.

\begin{figure}[htbp]
  \centering
  \includegraphics[width = 0.6\linewidth]{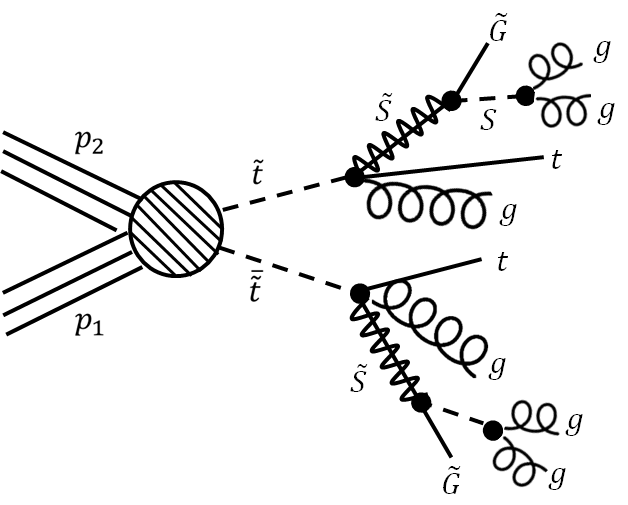}
  \caption{Top Squark production where $\tilde{t}$ decays to top quarks and gluons in the stealth $SY\bar{Y}$ model}
  \label{fig:top_squark}
\end{figure}

In 2021, the CMS collaboration published another stealth SUSY related analysis based on data collected at the LHC from 2016 - 2018 at $\unit[137]{fb^{-1}}$ integrated luminosity. The data were used to determine best-fit values and upper limits on the cross section for top squark pair production in the R-parity violation (RPV) SUSY and stealth SUSY model \cite{sirunyan2021search}. Figure~\ref{fig:top_squark} shows the stealth SUSY-relevant decay channel considered in this analysis. Here, each $\tilde{t}$ decays into a gluon, a top quark, and an $\tilde{S}$, with the $\tilde{S}$ subsequently decaying into an $S$ and a gravitino $\tilde{G}$. Finally, $S$ decays into jets via $S \to gg$. This paper is the first search for this decay chain at the LHC.

This analysis searched for new physics with experimental signatures of two top squarks plus many jets and low missing transverse momentum $(p^{miss}_T)$. The CMS collaboration was able to exclude top squark mass up to $\unit[870]{GeV}$ in the stealth SUSY scenario at the 95 $\%$ CL.

\begin{figure}[htbp]
    \centering
    \begin{minipage}{0.3\textwidth}
        \centering
        \includegraphics[width=\linewidth]{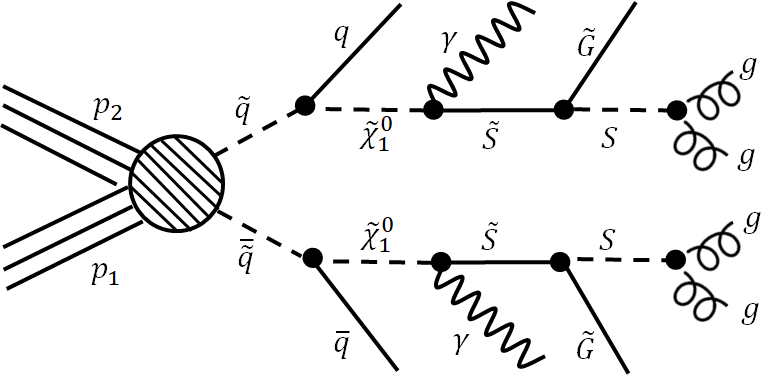}
    \end{minipage}\vspace{0.03\textwidth}
    \begin{minipage}{0.3\textwidth}
        \centering
        \includegraphics[width=\linewidth]{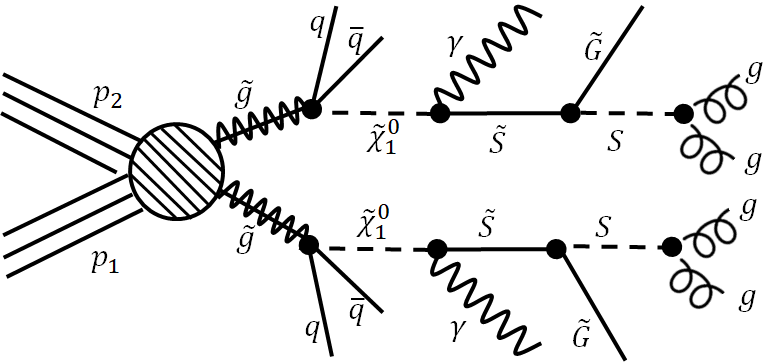}
    \end{minipage}
    \caption{Squark-pair (top) and gluino-pair (bottom) production in the stealth SUSY model. The final state consists of two photons, multiple jets, and low $p^{miss}_T$}
    \label{fig:cms2023}
\end{figure}

Finally, the most recent stealth SUSY result was published by the CMS Collaboration in 2023. The analysis used data collected by the CMS experiment in 2016 - 2018 at $\unit[138]{fb^{-1}}$ integrated luminosity \cite{cms2023search}. This work considered strongly produced SUSY particles where pairs of squarks or gluinos are produced in the primary interaction, as shown in Figure~\ref{fig:cms2023}. The SUSY particles decay into quarks and neutralinos $(\tilde{\chi}^0_1)$, with the neutralino subsequently decaying into the stealth sector $(\tilde{S})$ through the emission of photons.

This analysis specifically search for stealth SUSY in final states with two photons, at least four jets, and low missing transverse momentum in the final state. The analysis defined the quantity $S_T$ which is the scalar $p_T$ sum of all reconstructed particles in the event. This sum includes the momentum of jets, photons and missing transverse momentum $p^{miss}_T$.  The quantity $S_T$ was used to define the signal regions used in the analysis and each signal region was further subdivided into three jet multiplicities (4, 5, and $\geq$ 6). The CMS collaboration was able to exclude squark and gluino masses up to $\unit[1.85]{TeV}$ and $\unit[2.15]{TeV}$, respectively, at the 95 $\%$ CL within the framework of simplified stealth SUSY models. 

In this work, we will consider squark-pair production in the stealth SUSY model, the same decay chain shown in Figure~\ref{fig:cms2013}: 
\begin{equation}
\tilde{q}\to q(\tilde{\chi}^0_1\to \gamma (\tilde{S}\to \tilde{G}(S\to gg)))    
\end{equation}

This paper presents an estimated sensitivity for this production process in the stealth SUSY model, based on cuts that are independent of the missing energy. This also extends a phenomenological study that employs detector-related kinematic variables to reconstruct the bino resonance \cite{flores2020constraining}. The significance of the squark decay chain in this paper was evaluated after applying the kinematic cuts against the dominant QCD multijets background. The resulting sensitivity was then compared with the exclusion limits of the latest CMS stealth SUSY search \cite{cms2023search} and also to an ATLAS search that targets a different scenario but happens to be sensitive to the stealth SUSY model, as will be discusssed in Section \ref{sec:prev_searches}.

\section{Event Generation}
\label{sec:event_gen}

A total of 20,000 squark-pair production events were generated using \textsc{Pythia 8.245} \cite{sjostrand2015introduction} at $\sqrt{s} = \unit[13]{TeV}$ using the NNPDF 2.3 QCD + QED LO parton distribution function set \cite{ball2013_parton_qed}. A modified decay table was used so that the branching ratio is set to one for each decay chain in the process. The following set of masses were used: a fixed gluino mass $(m_{\tilde{g}})$ at $\unit[3000]{GeV}$, a squark mass $(m_{\tilde{q}})$ being varied from 1000 GeV to 2600 GeV and bino mass $(m_{\tilde{\chi}^0_1})$ also being varied from $\unit[150]{GeV}$ to $\unit[1950]{GeV}$; both are varied in steps of $\unit[50]{GeV}$. Lastly, the singlino and singlet masses are kept at $m_{\tilde{S}} = 100$ GeV and $m_S = 95$ GeV, respectively, with $\delta m = 5$ GeV. The squark pair-production cross section was calculated using \textsc{NNLLFast 2.0} \cite{beenakker2016nnll}. The total cross section of the signal ranges from $\unit[0.2504]{pb}$ for $m_{\tilde{q}} = \unit[1000]{GeV}$ to $\unit[0.00050152]{pb}$ for $m_{\tilde{q}} = \unit[2600]{GeV}$. 

For the Standard Model background, $5\times 10^6$ QCD multijets events with $p_T > \unit[300]{GeV}$ were also generated using \textsc{Pythia 8.245} with a cross section of $\unit[8.70\times 10^3]{pb}$. Other background processes, such as $t\bar{t}+Z$, $t\bar{t}+W$, $Z+\gamma$, and $W+\gamma$, were also evaluated, but were found to contribute negligibly after applying the selection criteria in a previous study \cite{flores2020constraining}. Consequently, only QCD multijets production was considered as a relevant background in this work.

\section{Other Searches at $\sqrt{s}= \unit[13]{TeV}$}
\label{sec:prev_searches}

To compare the results of the latest CMS stealth SUSY search with previous ATLAS and CMS analyses at $\sqrt{s} = \unit[13]{TeV}$, squark-pair decay chain events in the stealth SUSY model were simulated with \textsc{CheckMATE 2.0.41} for detector response and tested against previous search constraints. To determine whether a squark-bino mass point ($m _{\tilde{q}}, m _{\tilde{\chi}^0_1}$) is already excluded by a search or not, \textsc{CheckMATE} compares the estimate of signal events with the observed limits at 95\% CL using:
\begin{equation}
r = \frac{s-1.64\cdot \Delta s}{s^{95}_{exp}}
\end{equation}

\noindent
where $s^{95}_{exp}$ is the observed 95\% CL exclusion limit. The numerator corresponds to the 95\% CL lower bound on the prediction for the number of signal events. \textsc{CheckMATE} calculates the value of \textit{r} for every signal region of each analysis carried out at $\sqrt{s}=\unit[13]{TeV}$. A squark–bino mass point is excluded if at least one signal region within the analysis yields $r > 1$, and it is allowed if $r < 1$. 

\begin{figure}[htbp]
  \centering
  \includegraphics[width = 0.8\linewidth]{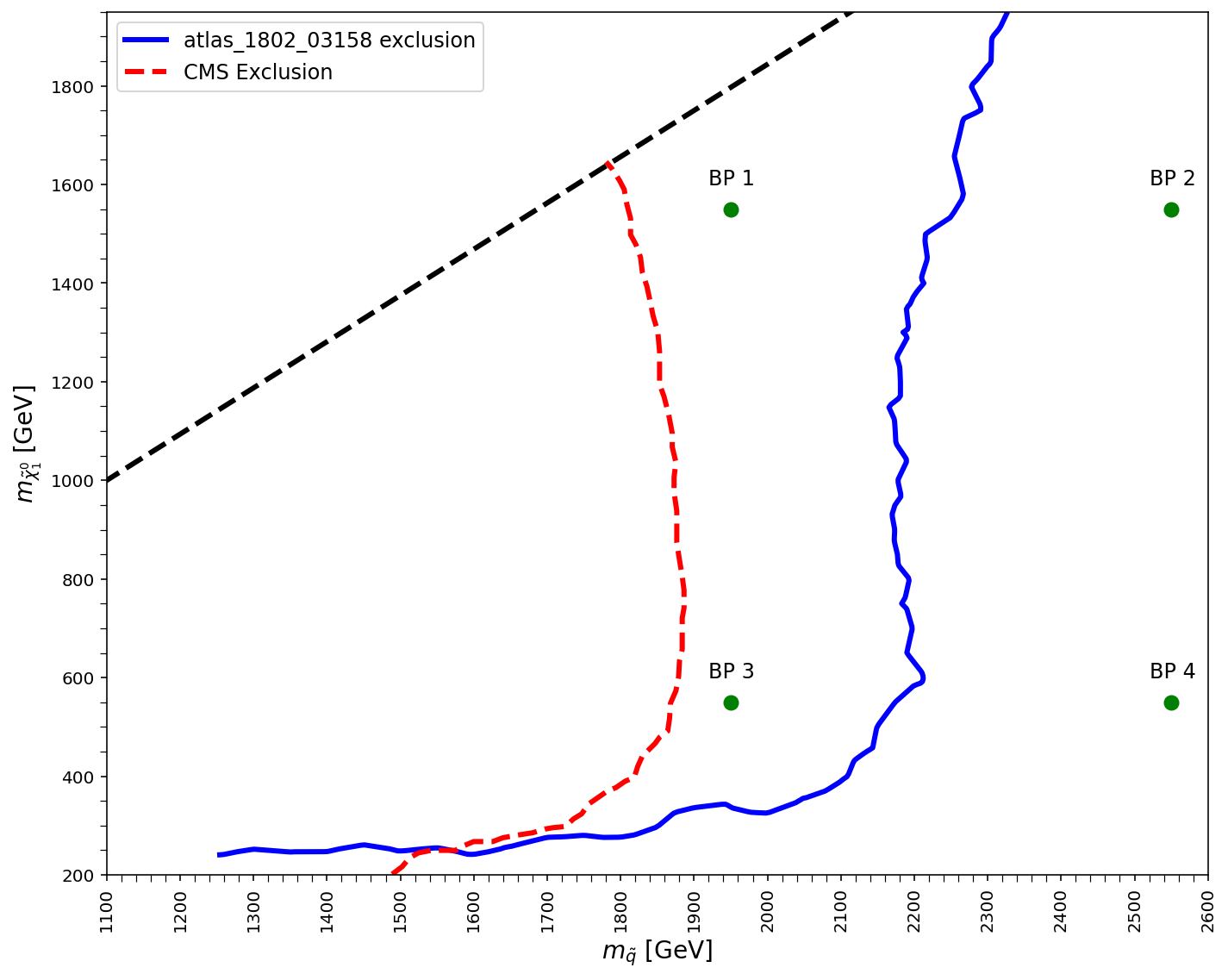}
  \caption{{\fontfamily{cmtt}\selectfont atlas\_1802\_03158} exclusion (blue) as recasted in \textsc{CheckMATE} vs CMS exclusion  for squark pair production in the stealth SUSY model (red) at 95 $\%$ CL exclusions.}
  \label{fig:squark_CheckMATE}
\end{figure}

After running \textsc{CheckMATE} for a squark mass ranging from 1000 - 2600 GeV and a bino mass ranging from 150 - 1950 GeV with focus on searches at $\sqrt{s}=\unit[13]{TeV}$, it was found out that the {\fontfamily{cmtt}\selectfont atlas\_1802\_03158} analysis provides a stronger exclusion compared to the CMS result as shown in Figure~\ref{fig:squark_CheckMATE}.

This analysis is an ATLAS search performed at $\sqrt{s}=\unit[13]{TeV}$ with an integrated luminosity of $\unit[36.1]{fb^{-1}}$, published in 2018 \cite{atlas2018_gmsb_photonic}. This search is motivated by generalized models of gauge-mediated supersymmetric breaking (GMSB) where the final states involve significant missing transverse momentum $(p^T_{miss})$ and photons. The solid blue line in Figure~\ref{fig:squark_CheckMATE} represents $r=1$ for the {\fontfamily{cmtt}\selectfont atlas\_1802\_03158} analysis according to \textsc{CheckMATE}, while the dashed red line is the CMS exclusion plot. The region to the left of the solid line is excluded ($r > 1$) while the region to the right is allowed ($ r < 1$). To investigate why the {\fontfamily{cmtt}\selectfont atlas\_1802\_03158} analysis made a better exclusion in the squark-bino mass phase space, four benchmark points (BP) were chosen with its corresponding \textsc{CheckMATE} result as summarized in Table~\ref{tab:checkMATE_benchmark}.

\begin{table}[htbp]
\centering
\caption{\textsc{CheckMATE} result of the {\fontfamily{cmtt}\selectfont atlas\_1802\_03158} analysis for different set of squark-bino pairs.}
\label{tab:checkMATE_benchmark}
\small  
\setlength{\tabcolsep}{4pt} 
\renewcommand{\arraystretch}{1.1} 
\begin{tabular}{|c|c|c|c|c|}
\hline
\textbf{BP} & $m_{\tilde{q}}$ (GeV) & $m_{\tilde{\chi}^0_1}$ (GeV) & \textbf{ \textsc{CheckMATE} result} & \textbf{r-value} \\
\hline
1 & 1750 & 1550 & Excluded & 5.58 \\
2 & 2550 & 1550  & Allowed  & 0.33 \\
3 & 1750 & 550  & Excluded & 2.69 \\
4 & 2550 & 550 & Allowed  & 0.35 \\
\hline 
\end{tabular}
\end{table}

The benchmark points can also be seen in  Figure~\ref{fig:squark_CheckMATE}. BPs 1 and 3 are excluded by the {\fontfamily{cmtt}\selectfont atlas\_1802\_03158} analysis while BPs 2 and 4 are allowed by the said analysis. For these benchmark points, the signal region $SR^{\gamma \gamma}_{S-L}$ of the {\fontfamily{cmtt}\selectfont atlas\_1802\_03158} analysis is the most sensitive signal region. It has a \textsc{CheckMATE} label {\fontfamily{cmtt}\selectfont SRaaSL} which means that this is a signal region [{\fontfamily{cmtt}\selectfont SR}] involving two photons [{\fontfamily{cmtt}\selectfont aa}] targeting the production of higher-mass strongly coupled SUSY states (gluino and squarks) [{\fontfamily{cmtt}\selectfont S}] optimized for lower-mass $\tilde{\chi}^0_1$ [{\fontfamily{cmtt}\selectfont L}]. The selection criteria for this signal region is summarized in Table~\ref{tab:checkMATE_SRaaSL} \cite{atlas2018_gmsb_photonic}.

\begin{table}[htbp]
\centering
\caption{Selection criteria for signal region $SR^{\gamma \gamma}_{S\text{-}L}$ of the {\fontfamily{cmtt}\selectfont atlas\_1802\_03158} analysis}
\label{tab:checkMATE_SRaaSL}
\begin{tabular}{l}
\hline
\multicolumn{1}{l}{\textbf{Selection Criteria}} \\ \hline
\multicolumn{1}{c}{Number of photons $\geq$ 2}                             \\
\multicolumn{1}{c}{$E^\gamma_T$ [GeV] $>$ 75}                             \\
\multicolumn{1}{c}{$\slashed{E}_T$ [GeV] $>$ 150}                             \\
\multicolumn{1}{c}{$H_T$ [GeV] $>$ 2750}                             \\
\multicolumn{1}{c}{$\Delta\phi_{\text{min}}(\text{jet}, \slashed{E}_T) > 0.5$} \\ \hline
\end{tabular}
\end{table}

\noindent
Here, $E^\gamma_T$ is the energy requirement for tight isolated photons, $\slashed{E}_T$ is the magnitude of the missing transverse momentum, $H_T$ is defined as the transverse energy of photons and additional photons and jets in the event without the addition of $\slashed{E}_T$, and $\Delta\phi_{\text{min}}(\text{jet}, \slashed{E}_T)$ is the azimuthal angle between the jet candidate and $\slashed{E}_T$ observable. In the result of the {\fontfamily{cmtt}\selectfont atlas\_1802\_03158} analysis, they were able to observe $S^{95}_{\text{obs}} = 3.0$ events at 95$\%$ CL for the $SR^{\gamma \gamma}_{S\text{-}L}$ signal region. This information allows us to exclude a squark-bino mass pair if it has more than 3 remaining events after applying the selection criteria defined for the {\fontfamily{cmtt}\selectfont SRaaSL} signal region.

\begin{figure}[htbp]
  \centering
  \begin{minipage}{0.8\linewidth}
    \centering
    \includegraphics[width=\linewidth]{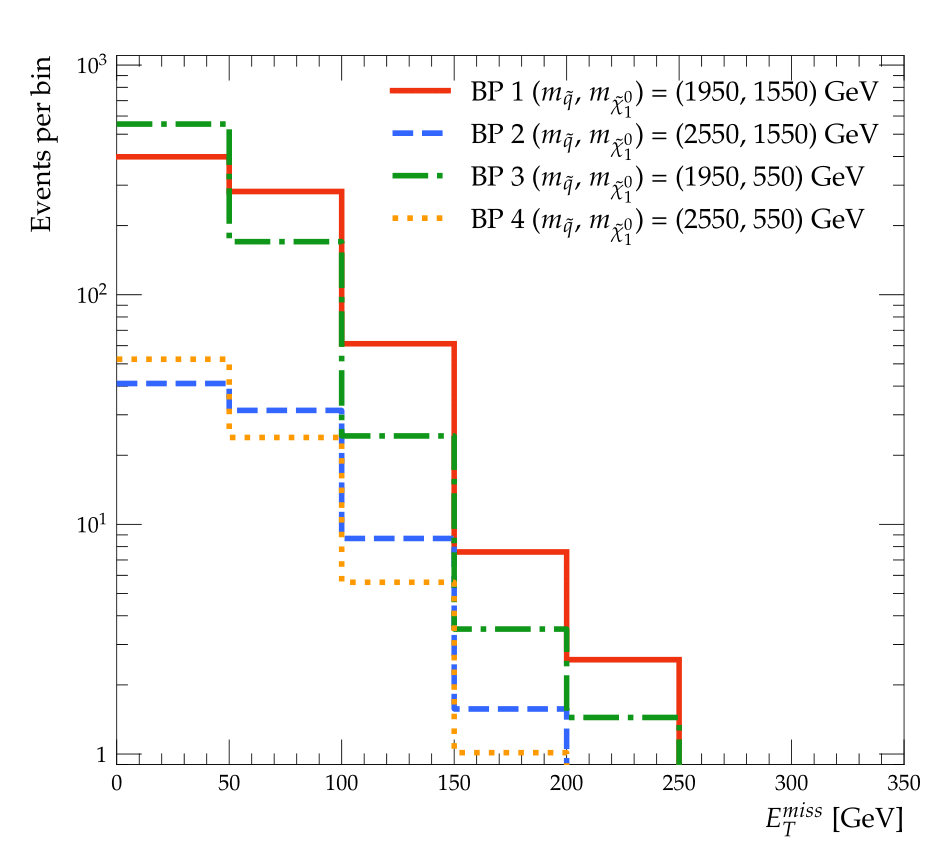}
    \label{fig:ET_miss_all}
  \end{minipage}
  \hfill
  \begin{minipage}{0.8\linewidth}
    \centering
    \includegraphics[width=\linewidth]{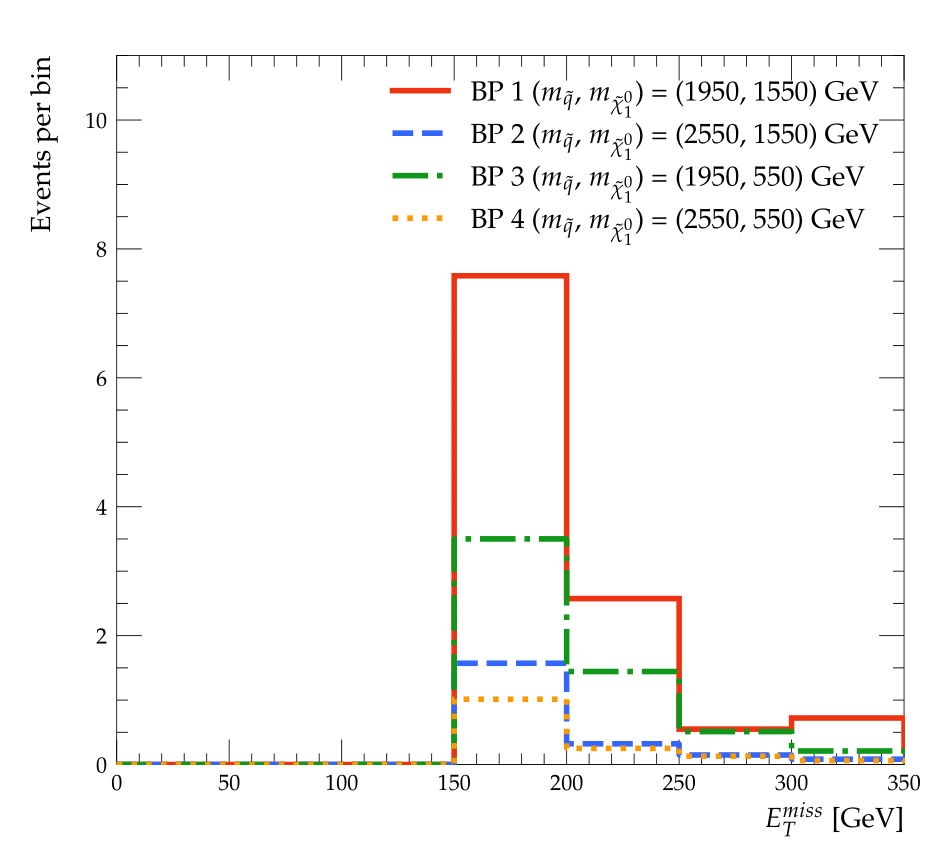}
    \label{fig:ET_miss_all_150}
  \end{minipage}\
  \caption{$\slashed{E}_T$ distribution of the benchmark points before (top) and after (bottom) the $\slashed{E}_T >\unit[150]{GeV}$ cut.}
  \label{fig:ET_miss}
\end{figure}

Figure~\ref{fig:ET_miss} shows the $\slashed{E}_T$ distribution of the chosen benchmark points before and after applying the $\slashed{E}_T >\unit[150]{GeV}$ cut. The remaining events were counted per benchmark point and summarized in Table~\ref{tab:checkMATE_ETmiss}.

\begin{table}[htbp]
\centering
\caption{Remaining events per BP after the $\slashed{E}_T >\unit[150]{GeV}$ cut and the remaining signal count after implementing the $SR^{\gamma \gamma}_{S\text{-}L}$ selection criteria}
\label{tab:checkMATE_ETmiss}
\small  
\setlength{\tabcolsep}{4pt} 
\renewcommand{\arraystretch}{1.1} 
\begin{tabular}{|c|c|c|c|c|}
\hline
\textbf{BP} & $m_{\tilde{q}}$ (GeV) & $m_{\tilde{\chi}^0_1}$ (GeV) & \textbf{$\slashed{E}_T$ miss cut} & $SR^{\gamma \gamma}_{S\text{-}L}$ (s) \\
\hline
1 & 1950 & 1550 & 12.7 & 8.1 \\
2 & 2550 & 550  & 2.3 &  1.1  \\
3 & 1950 & 550  & 6.1 & 6.0 \\
4 & 2550 & 1550 & 1.6 & 1.1  \\
\hline 
\end{tabular}
\end{table}

The {\fontfamily{cmtt}\selectfont atlas\_1802\_03158} analysis looks for experimental signatures that incorporate an isolated photon and a significant amount of $\slashed{E}_T$. As seen in BPs 1 and 3, a significant number of events are still present in the $\slashed{E}_T$ distribution after implementing the $\slashed{E}_T >\unit[150]{GeV}$ cut. This does not hold for BPs 2 and 4, with only 2.3 and 1.6 events remaining after the $\slashed{E}_T$ cut. Thus, the region of the squark-bino mass phase space beyond the CMS exclusion, where BP1 and BP3 are located, becomes sensitive to the {\fontfamily{cmtt}\selectfont atlas\_1802\_03158} analysis due to the large number of $\slashed{E}_T$ events remaining after implementing the selection criteria defined in the $SR^{\gamma \gamma}_{S-L}$ signal region. In terms of stealth SUSY, increasing the bino mass also increases the energy being carried off by the gravitino $\tilde{G}$, which is the LSP and carrier of $\slashed{E}_T$ in the final state of this model.

The $\slashed{E}_T$ distributions of the selected benchmark points illustrate why the {\fontfamily{cmtt}\selectfont atlas\_1802\_03158} analysis achieves a stronger exclusion than the latest CMS stealth SUSY search. In the next section, we try to extend the sensitivity of squark-pair production in the stealth SUSY model for regions with a small amount of $\slashed{E}_T$ events and thus evades the {\fontfamily{cmtt}\selectfont atlas\_1802\_03158} and the most recent CMS analysis targetting stealth SUSY.

\section{Signal and Background Events Analysis}
\label{sec:analysis}

The analysis of the stealth SUSY signal and the QCD multijets background events was implemented using the \textsc{Rivet 4.0.1} analysis toolkit \cite{bierlich2020robust}. The requirements for jets and photons are the same as the conditions used in a previous stealth SUSY study \cite{flores2020constraining}. The jets were clustered using the FasTJet 3.4.3 anti-kT algorithm \cite{cacciari2012fastjet} and the large-radius ($R = 1.0$) jets were trimmed \cite{krohn2010jet} with requirements of $p_T > \unit[450]{GeV}$ and $|\eta| < 1.5$. Photon candidates were required to satisfy the condition that $p_T > \unit[200]{GeV}$ and $|\eta| < 2.0$. A lepton veto was also included in the final state. The analysis involves kinematic variables related to the decay products and quantities measured in the detector ($p_T$, $\Delta R_{ij}$, $\Delta \eta$, multiplicity, etc.). Finally, the signal and background histograms for each variable were normalized to an integrated luminosity of $\unit[138]{fb^{-1}}$, matching that of the recent CMS stealth SUSY search. To compare the performance of these cuts, the signal significance ($\sigma$) was calculated using:
\begin{equation}
\sigma = \frac{S}{\sqrt{S+B}}
\end{equation}  

\noindent
where $S$ denotes the remaining stealth SUSY signal events and $B$ is the remaining QCD multijets background events after applying the selection criteria. A significance value of at least $3\sigma$ indicates that a corresponding squark–bino mass pair ($m_{\tilde{q}}$ / $m_{\tilde{\chi}^0_1}$) would be within the potential discovery reach of future stealth SUSY searches.

\section{Definition of Signal Regions}
\label{sec:signal_region}  

In this paper, we defined two signal regions with their corresponding selection criteria. We determined that these kinematic cuts improved the sensitivity in relation to the QCD multijets background.  

\subsection{Signal Region I}

From a previous stealth SUSY study \cite{flores2020constraining}, a selection criteria was used that involved the multiplicity of large-radius jets in an event. In this paper, the said criteria was modified to include the angular separation between the leading and next-leading photons in an event $(\Delta R_{\gamma_1, \gamma_2})$. We define the first signal region ({\fontfamily{cmtt}\selectfont SR I}) with the selection criteria as follows:

\begin{enumerate}
    \item The number of leading-mass large-radius jets ($R = 1.0$) in an event must be greater than 3.
    \item The angular distance between the leading and the next-leading photons $(\Delta R_{\gamma_1, \gamma_2})$ in an event must be greater than 0.10.
\end{enumerate}

\begin{figure}[htbp]
  \centering
  \includegraphics[width = 0.8\linewidth]{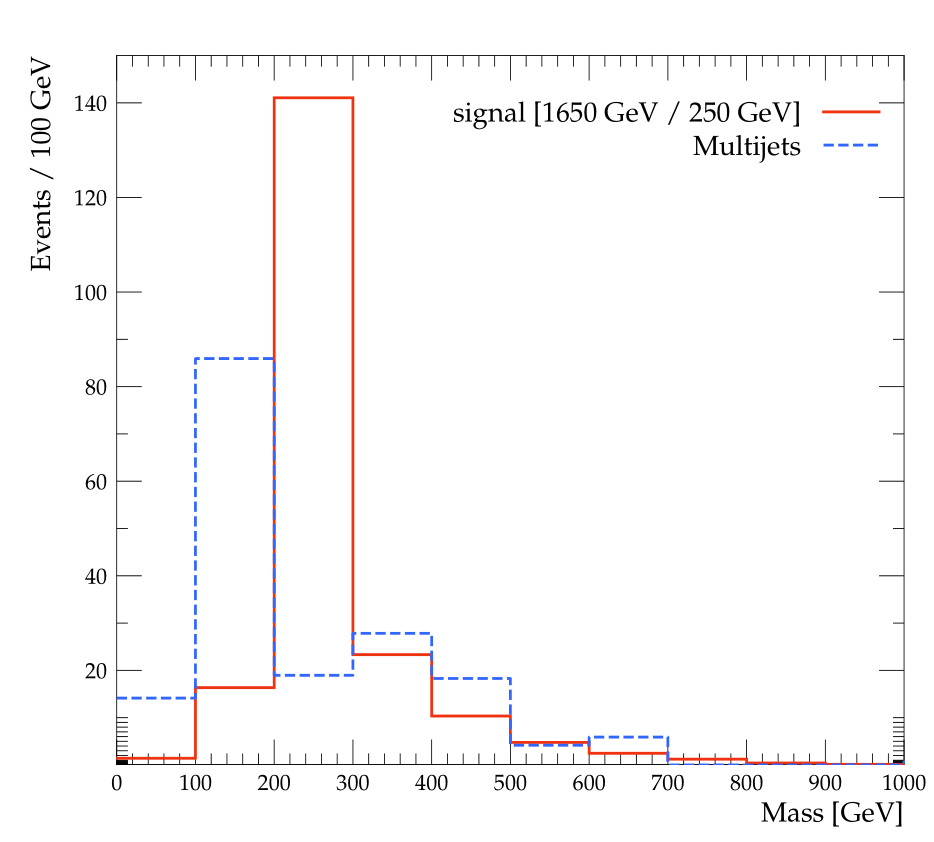}
  \caption{Mass distribution of the leading-mass large-radius jet for signal point $(m_{\tilde{q}}, m_{\tilde{\chi}^0_1}) = (\unit[1650]{GeV}, \unit[250]{GeV})$ together with the QCD multijets background in the 200-300 GeV region. The events are normalized using an integrated luminosity of $\unit[138]{fb^{-1}}$.}
  \label{fig:sample_SR1}
\end{figure}

A sample signal point in the squark-bino mass phase space at $(m_{\tilde{q}}, m_{\tilde{\chi}^0_1}) = (\unit[1650]{GeV}, \unit[250]{GeV})$ was selected. The selection criteria in {\fontfamily{cmtt}\selectfont SR I} were applied and the mass distribution of the leading-mass large-radius jets for this point was plotted against the QCD multijets background as shown in Figure~\ref{fig:sample_SR1}. The number of leading-mass large-radius jet in the 200 - 300 GeV range was investigated since the signal peaks at $\unit[250]{GeV}$ for this sample signal point. It is clearly shown that the number of signal events were able to overcome the number of QCD multijets background events in this range.

Table~\ref{tab:cutflow_SR1} shows the remaining events for the sample signal point ( $m_{\tilde{q}} = \unit[1650]{GeV}$ / $m_{\tilde{\chi}^0_1} = \unit[250]{GeV}$) vs. QCD multijets background events after the {\fontfamily{cmtt}\selectfont SR I} kinematic cuts. Using equation (5), the sensitivity for this point was calculated to be around 11.2 at $\unit[138]{fb^{-1}}$ integrated luminosity. 

\begin{table}[htbp]
\centering
\caption{Cutflow table showing the number of remaining events for the signal point $(m_{\tilde{q}}, m_{\tilde{\chi}^0_1}) = (\unit[1650]{GeV}, \unit[250]{GeV})$ and QCD multijets in the 200-300 GeV region. The events are normalized using an integrated luminosity of $\unit[138]{fb^{-1}}$.}
\label{tab:cutflow_SR1}
\begin{tabular}{|c|c|c|}
\hline
\multirow{2}{*}{} & \textbf{Signal} & \textbf{multijets} \\
                                         & $\unit[0.0166]{pb}$ & $\unit[8.70\times 10^3]{pb}$\\ \hline
No cuts                                  & 946   & $7.58 \times 10^7$   \\ 
Large Jet Multiplicity $> 3$             & 251   & $7.94 \times 10^3$  \\ 
$\Delta R_{\gamma_1,\gamma_2} > 0.1$           & 141 & 19    \\ \hline
\end{tabular}
\end{table}

\begin{figure}[htbp]
  \centering
  \includegraphics[width = 0.7\linewidth]{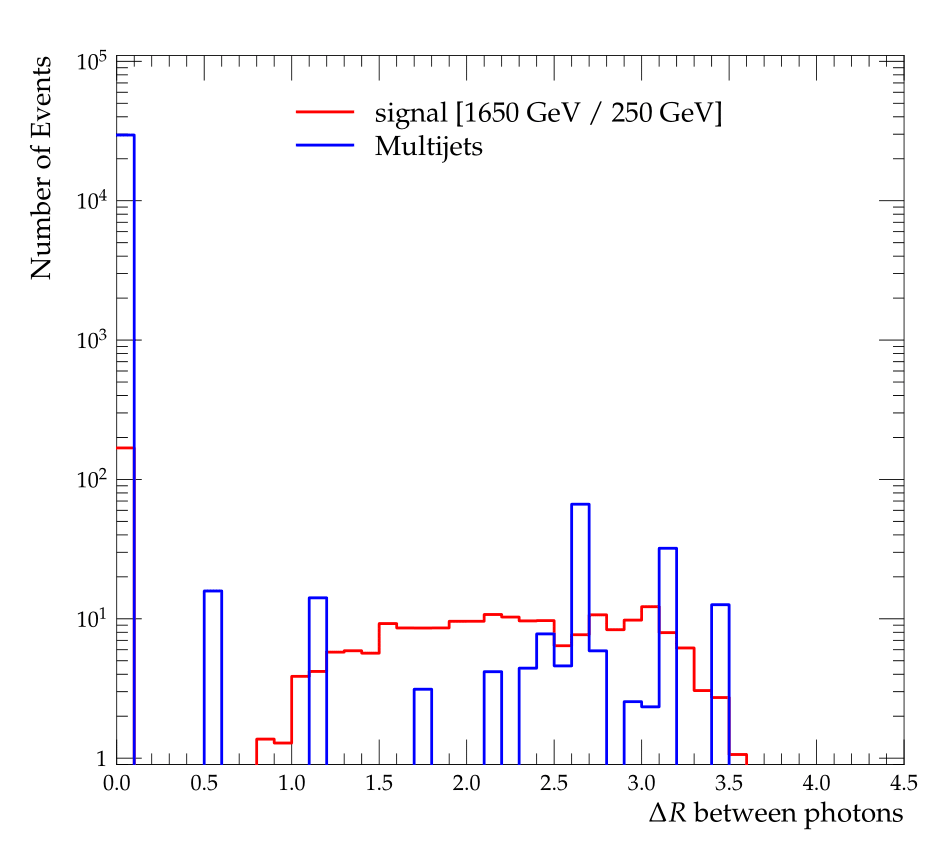}
  \caption{$\Delta R_{\gamma_1, \gamma_2}$ distribution for $(m_{\tilde{q}}, m_{\tilde{\chi}^0_1}) = (\unit[1650]{GeV}, \unit[250]{GeV})$ vs QCD multijets background after the large-radius jet multiplicity $>$ 3 cut}
  \label{fig:deltaR}
\end{figure}

\begin{figure}[htbp]
  \centering
  \includegraphics[width = 0.6\linewidth]{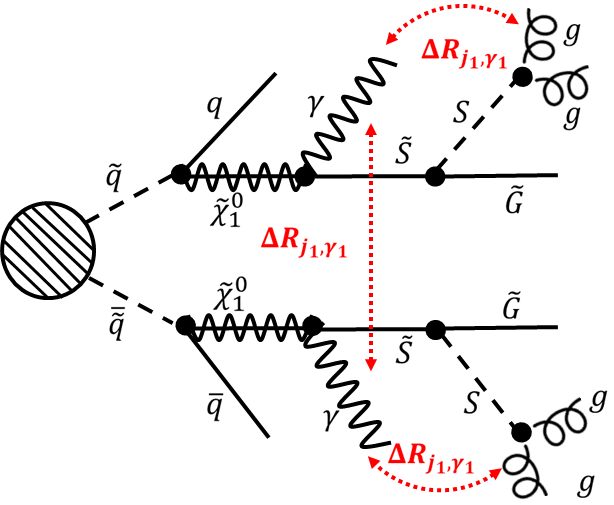}
  \caption{$\Delta R$ between final states of the squark-pair production in the stealth SUSY model}
  \label{fig:diagram_deltaR_new}
\end{figure}

Figure~\ref{fig:deltaR} shows the $\Delta R_{\gamma_1,\gamma_2}$ distribution for the sample signal point $(m_{\tilde{q}}, m_{\tilde{\chi}^0_1}) = (\unit[1650]{GeV}, \unit[250]{GeV})$ and the QCD multijet background, after requiring more than three large-radius jets per event. As shown in the figure, majority of QCD multijets events peak near 0 while events for the sample signal point are distributed from 0 to at most 3.6. A large angular separation is expected, as the leading and next-leading photons come from separate decay chains, as shown in Figure~\ref{fig:diagram_deltaR_new}. Their directions are therefore uncorrelated, resulting in a substantial angular separation between the two photons \cite{ghosh2010_low_missing_lookalikes}. QCD multijets do not have this feature, so the $\Delta R_{\gamma_1,\gamma_2}$ distribution between its leading and next-leading photons is almost uniform. Thus, the $\Delta R_{\gamma_1,\gamma_2} > 0.10$ cut was utilized to greatly reduce the number of background events in the dataset. 

The selection criteria defined by {\fontfamily{cmtt}\selectfont SR I} were also applied to the squark and bino mass ranges discussed in Section~\ref{sec:event_gen}, and the resulting sensitivity is presented in the next subsection.

\subsubsection{Sensitivity Region for Signal Region I}
\label{subsec:sri}

\begin{figure}[htbp]
  \centering
  \includegraphics[width = 0.8\linewidth]{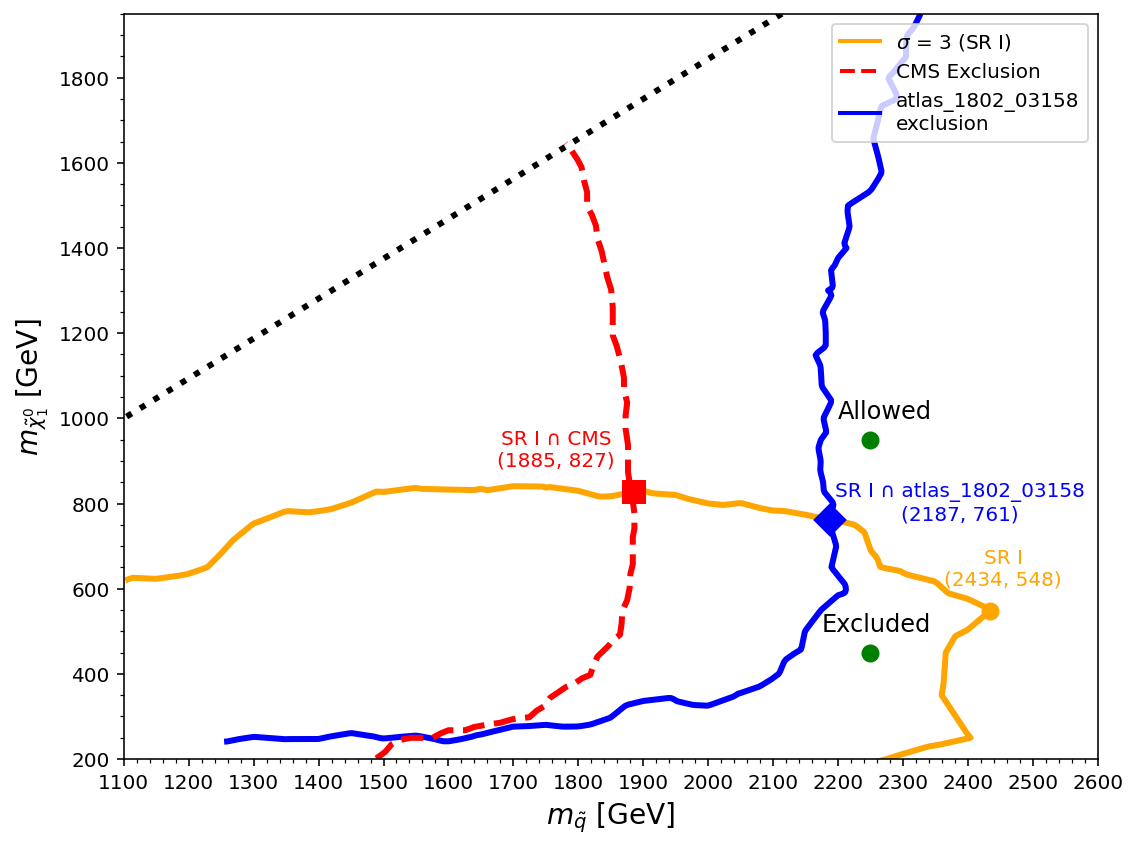}
  \caption{Comparison of the {\fontfamily{cmtt}\selectfont SR I} sensitivity vs CMS exclusion. The events are normalized using an integrated luminosity of $\unit[138]{fb^{-1}}$.}
  \label{fig:cms_SRI}
\end{figure}

The solid orange line shown in Figure~\ref{fig:cms_SRI} represents the extended sensitivity in the squark-bino mass phase space, obtained by applying the {\fontfamily{cmtt}\selectfont SR I} selection criteria. This extension in sensitivity is compared with the dashed red line, representing the exclusion from the latest CMS stealth SUSY search of the same decay chain \cite{cms2023search}, and with the solid blue line, representing the {\fontfamily{cmtt}\selectfont atlas\_1802\_03158} analysis exclusion obtained from $\sqrt{s}=\unit[13]{TeV}$ searches using \textsc{CheckMATE}. 

The region above the black diagonal line is a forbidden region based on the values of the parameters  $(m_{\tilde{q}}, m_{\tilde{\chi}^0_1}, m_{\tilde{S}}, m_S)$ set in Section~\ref{sec:event_gen}. The {\fontfamily{cmtt}\selectfont SR I} was able to extend the sensitivity beyond the CMS and ATLAS exclusion up to about $\unit[2400]{GeV}$ squark mass for a bino mass of around $\unit[550]{GeV}$. For comparison, the CMS analysis excludes squark masses up to $\unit[1850]{GeV}$ for the same decay chain.

However, for bino masses above $\unit[750]{GeV}$, the sensitivity of {\fontfamily{cmtt}\selectfont SR I} is significantly reduced. Above the point $(m_{\tilde{q}}, m_{\tilde{\chi}^0_1}) = (\unit[1885]{GeV}, \unit[827]{GeV})$, the CMS analysis yields a stronger exclusion in the squark-bino mass phase space. Also, the {\fontfamily{cmtt}\selectfont atlas\_1802\_03158} analysis yields a stronger exclusion above the $(m_{\tilde{q}}, m_{\tilde{\chi}^0_1}) = (\unit[2187]{GeV}, \unit[761]{GeV})$ point. To analyze this result, two sample points labeled ``Excluded" and ``Allowed" were chosen. The excluded point of {\fontfamily{cmtt}\selectfont SR I} has a squark-bino mass $(m_{\tilde{q}}, m_{\tilde{\chi}^0_1}) = (\unit[2250]{GeV}, \unit[450]{GeV})$ while the allowed point has a squark-bino mass $(m_{\tilde{q}}, m_{\tilde{\chi}^0_1}) = (\unit[2250]{GeV}, \unit[950]{GeV})$, respectively. 

\begin{figure}[htbp]
  \centering
  \includegraphics[width = 0.9\linewidth]{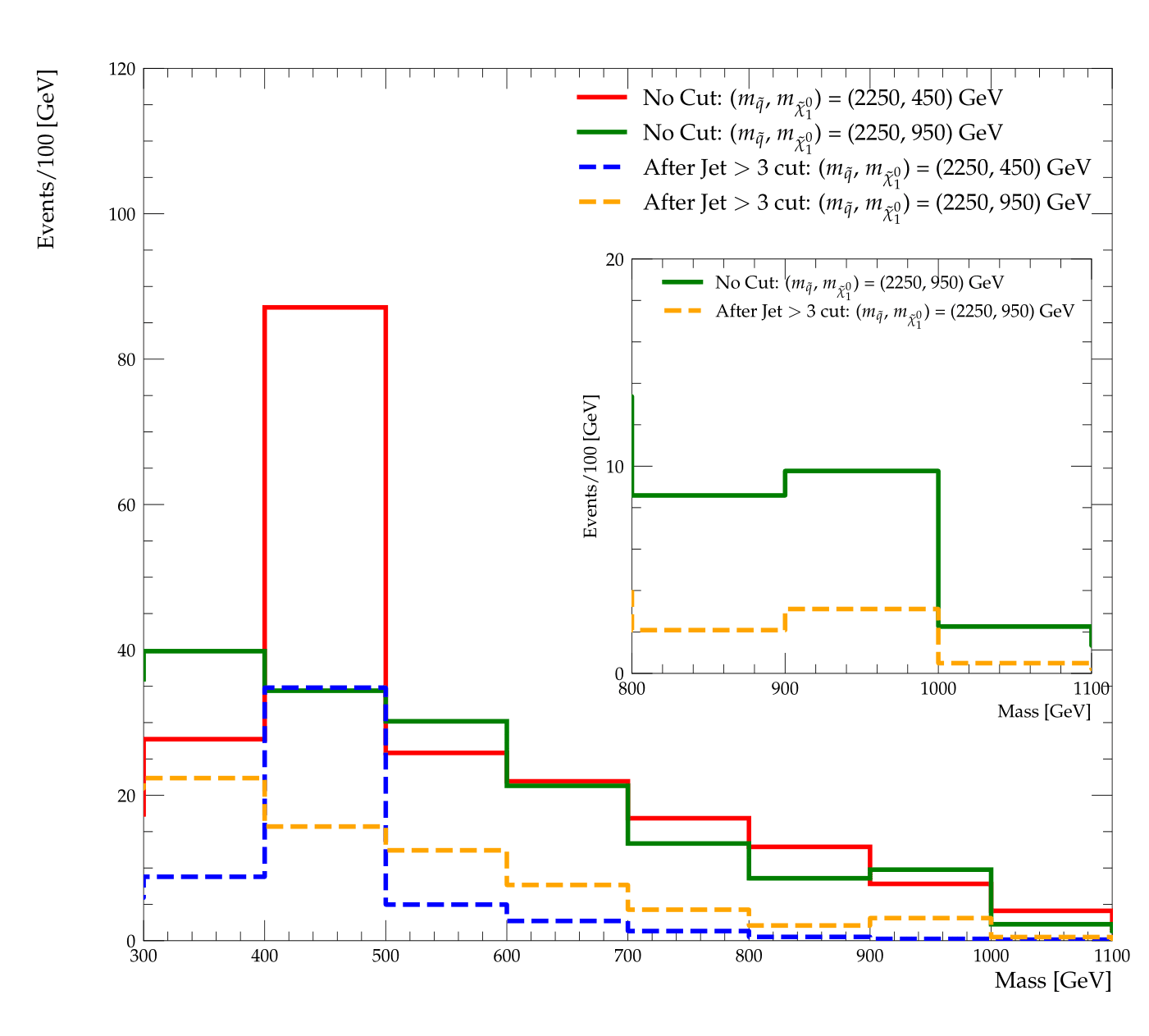}
  \caption{Mass distribution of leading-mass large-radius jet for the excluded and allowed point of {\fontfamily{cmtt}\selectfont SR I}}
  \label{fig:jet3_cut}
\end{figure}

The mass distributions of the leading large-mass large-radius jet of the two points were plotted and compared with each other as shown in Figure~\ref{fig:jet3_cut}. As seen in the figure, the large-radius jet multiplicity $>$ 3 cut greatly affected the point with a higher bino mass $(m_{\tilde{q}}=\unit[950]{GeV})$ compared to the point with a lower bino mass $(m_{\tilde{q}}=\unit[450]{GeV})$. This is also highlighted in Table~\ref{tab:compare_jet3}, where the heavier bino mass lost approximately $68.19 \%$ of the signal after this particular cut compared to $60.04 \%$ for the lighter bino mass.

Specifically, we counted events in the 400–500 GeV range for the “Excluded” point and in the 900–1000 GeV range for the “Allowed” point, corresponding to the regions where the signal peaks for these points.

\begin{table}[htbp]
\centering
\caption{Remaining signal events for the excluded and allowed point after the large-radius jet multiplicity $>$ 3 cut. The events are normalized using an integrated luminosity of $\unit[138]{fb^{-1}}$.}
\label{tab:compare_jet3}
\begin{tabular}{|c|c|c|}
\hline
\multirow{2}{*}{$m_{\tilde{q}} = \unit[2250]{GeV}$} & \textbf{Excluded} & \textbf{Allowed} \\
                                         & $m_{\tilde{\chi}^0_1}=\unit[450]{GeV}$ & $m_{\tilde{\chi}^0_1}=\unit[950]{GeV}$\\ \hline
No cuts                                  & 87.1   & 9.8   \\ 
Large Jet Multiplicity $> 3$             & 34.8   & 3.1  \\ \hline
\textbf{$\%$ of signal loss }            & \textbf{60.04}   & \textbf{68.19}  \\ \hline
\end{tabular}
\end{table}

This behavior is expected, as a highly energetic particle produces decay products that appear collimated in the lab frame due to its large Lorentz boost \cite{lapsien2016new}. For a bino with heavier mass, the singlino is boosted and the decay products in the final state will be collimated. The jet algorithm, in turn, will cluster them into a single jet due to the small angular separation between the decay products. This is not the case for a signal with lighter bino mass, which still has a lot of events remaining after the large-radius jet multiplicity cut. For a lighter bino, its decay products are more spread out in the lab frame, allowing for better jet resolution and separation.

Thus, requiring a large-radius jet multiplicity greater than three substantially reduces the number of signal events for $(>\unit[750]{GeV})$ bino mass, leading to a significance that is too low to be considered sensitive in this region. Motivated by this observation, we define an additional signal region to extend the sensitivity to bino masses above $\unit[750]{GeV}$ and beyond the {\fontfamily{cmtt}\selectfont atlas\_1802\_03158} analysis exclusion.

\subsection{Signal Region II}

Since the requirement of more than three large-radius jets in an event as required by {\fontfamily{cmtt}\selectfont SR I} reduces the sensitivity for squark–bino mass points for $>\unit[750]{GeV}$ bino masses, it is necessary to relax this cut to extend the sensitivity beyond this bino massz. Consequently, for the next signal region ({\fontfamily{cmtt}\selectfont SR II}), we define the selection criteria as follows:

\begin{enumerate}
    \item The number of leading-mass large-radius jets ($R = 1.0$) in an event must be \textbf{greater than 2}.
    \item The angular distance between leading and the next-leading photons $(\Delta R_{\gamma_1, \gamma_2})$ in an event must be greater than 0.10.
    \item The angular distance between leading large jet and leading photon $(\Delta R_{j_1, \gamma_1})$ in an event must be greater than 0.20.
\end{enumerate}

\begin{figure}[htbp]
  \centering
  \includegraphics[width = 0.8\linewidth]{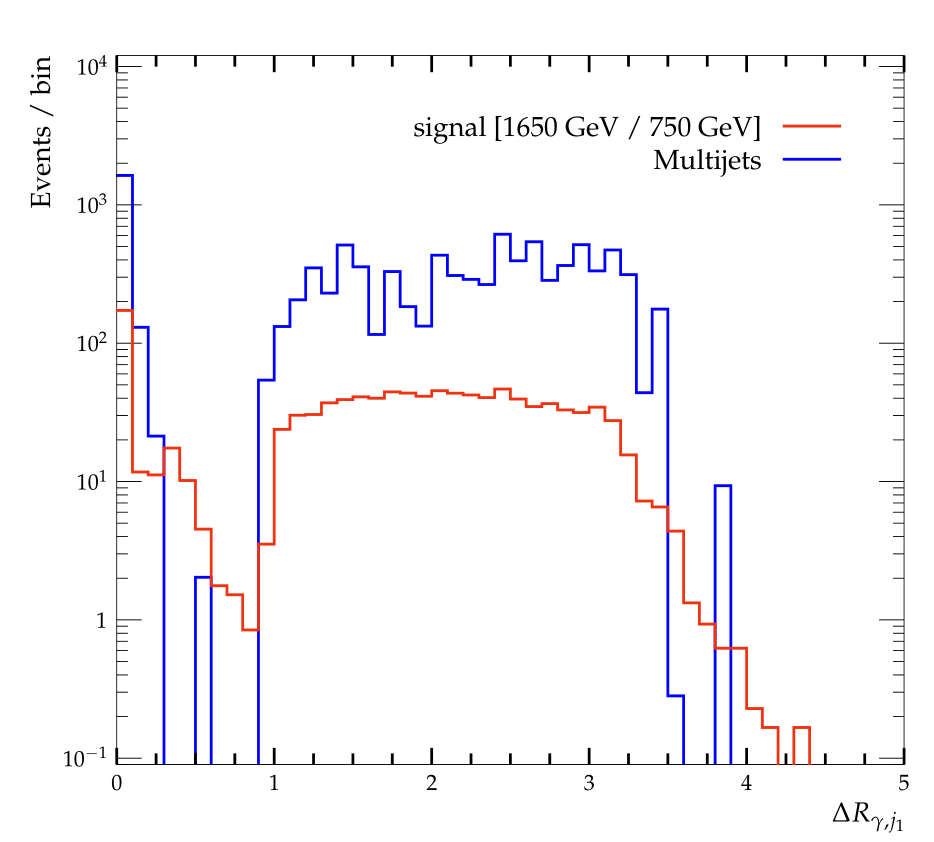}
  \caption{$\Delta R_{j_1, \gamma_1}$ distribution for a sample signal point $(m_{\tilde{q}}, m_{\tilde{\chi}^0_1}) = (\unit[1650]{GeV}, \unit[750]{GeV})$ and QCD multijets background after the $\Delta R_{\gamma_1, \gamma_2} >0.1$ cut.  The events are normalized using an integrated luminosity of $\unit[138]{fb^{-1}}$.}
  \label{fig:deltaR_SR2}
\end{figure}

Figure~\ref{fig:deltaR_SR2} shows the $\Delta R_{j_1, \gamma_1}$ comparison between the QCD multijets background (blue) and a sample signal point $(m_{\tilde{q}}, m_{\tilde{\chi}^0_1}) = (\unit[1650]{GeV}, \unit[750]{GeV})$ (red) after the first two selection criteria described above. As seen in the figure, QCD multijets events are highly populated in the region $\Delta R_{\gamma_1, j_1} \leq 0.2$. In QCD multijets, most photons originate from quark bremsstrahlung or parton fragmentation. In these processes, high-energy partons (quarks or gluons) hadronize into observed jets, and the resulting fragmentation photons are emitted collinearly with the initiating parton. This explains the small angular separation between jets and photons for this background $(\Delta R_{j_1, \gamma})$. In contrast, jets and photons for the stealth SUSY signal arise from the decay of a heavy bino, and the decay kinematics produce a large angular separation between the final-state particles which can also be seen in Figure~\ref{fig:diagram_deltaR_new}. After imposing the last selection criterion in {\fontfamily{cmtt}\selectfont SR II}, the number of QCD multijets events was reduced by around 1762 while only affecting 184 signal events for this sample signal point. 

\noindent
Table~\ref{tab:cutflow_SR2} shows the remaining events for the sample signal point ( $m_{\tilde{q}} = \unit[1650]{GeV}$ / $m_{\tilde{\chi}^0_1} = \unit[750]{GeV}$) vs. QCD multijets background after imposing the selection criteria defined in {\fontfamily{cmtt}\selectfont SR II}. We look for the number of leading-mass large-radius jets in the 700 - 800 GeV range since the signal peaks around $\unit[750]{GeV}$ for this point.

\begin{table}[htbp]
\centering
\caption{Cutflow table showing the number of events for a sample  signal point $(m_{\tilde{q}}, m_{\tilde{\chi}^0_1}) = (\unit[1650]{GeV}, \unit[750]{GeV})$ and QCD multijets background at 700-800 GeV region for {\fontfamily{cmtt}\selectfont SR II} selection criteria. The events are normalized using an integrated luminosity of $\unit[138]{fb^{-1}}$.}
\label{tab:cutflow_SR2}
\begin{tabular}{|c|c|c|}
\hline
\multirow{2}{*}{} & \textbf{Signal} & \textbf{multijets} \\
                                         & $\unit[0.0166]{pb}$ & $\unit[8.70\times 10^3]{pb}$\\ \hline
No cuts                                  & 109.3   & $2.40 \times 10^4$   \\ 
Large Jet Multiplicity $> 2$             & 79.7  & $2.23 \times 10^3$  \\ 
$\Delta R_{\gamma_1,\gamma_2} > 0.1$     & 66.4 & 26.4   \\ 
$\Delta R_{\gamma_1,j_1} > 0.2$          & 59.2 & 3.5    \\ \hline
\end{tabular}
\end{table}

Using equation (5) for this signal point, the sensitivity was calculated to be around 7.5 at $\unit[138]{fb^{-1}}$ integreated luminosity. If we apply the selection criteria defined in {\fontfamily{cmtt}\selectfont SR I} for the same signal point, we get the following number of remaining events listed in Table~\ref{tab:cutflow_SR1_v2}:

\begin{table}[htbp]
\centering
\caption{Cutflow table showing the number of events for a sample signal point $(m_{\tilde{q}}, m_{\tilde{\chi}^0_1}) = (\unit[1650]{GeV}, \unit[750]{GeV})$ and QCD multijets background at 700-800 GeV region for {\fontfamily{cmtt}\selectfont SR I} selection criteria. The events are normalized using an integrated luminosity of $\unit[138]{fb^{-1}}$.}
\label{tab:cutflow_SR1_v2}
\begin{tabular}{|c|c|c|}
\hline
\multirow{2}{*}{} & \textbf{Signal} & \textbf{multijets} \\
                                         & $\unit[0.0166]{pb}$ & $\unit[8.70\times 10^3]{pb}$\\ \hline
No cuts                                  & 109.3   & $2.40 \times 10^4$   \\ 
Large Jet Multiplicity $> 3$             & 20.6  & 38.5  \\ 
$\Delta R_{\gamma_1,\gamma_2} > 0.1$     & 18.1 & 0   \\ \hline
\end{tabular}
\end{table}

\begin{figure}[htbp]
  \centering
  \includegraphics[width = 0.7\linewidth]{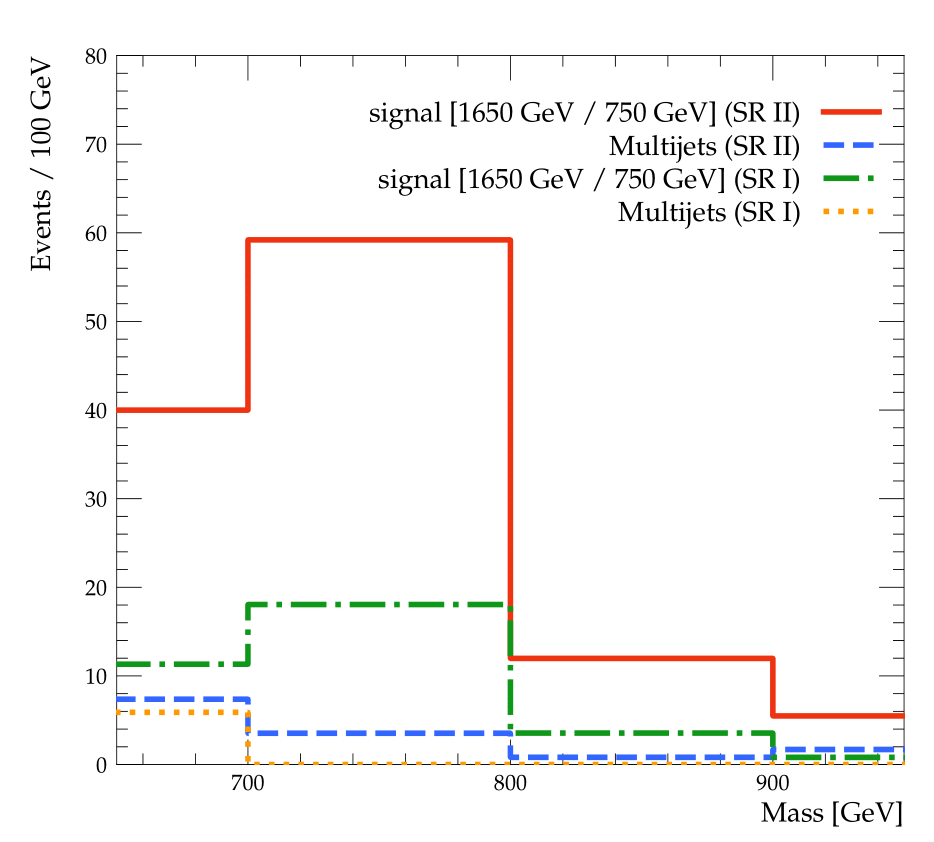}
  \caption{Mass distribution of the leading-mass large-radius jet for a sample signal point $(m_{\tilde{q}}, m_{\tilde{\chi}^0_1}) = (\unit[1650]{GeV}, \unit[750]{GeV})$ together with QCD multijets background at 700-800 GeV region. The events are normalized using an integrated luminosity of $\unit[138]{fb^{-1}}$.}
  \label{fig:comp_SR1_SR2}
\end{figure}

Using equation (5) will yield a sensitivity of around 4.2 at $\unit[138]{fb^{-1}}$ integrated luminosity, which is lower than the {\fontfamily{cmtt}\selectfont SR II} result for the same signal point.  The mass distributions of the leading-mass large-radius jets of the signal point $(m_{\tilde{q}}, m_{\tilde{\chi}^0_1}) = (\unit[1650]{GeV}, \unit[750]{GeV})$ upon application of {\fontfamily{cmtt}\selectfont SR I} and {\fontfamily{cmtt}\selectfont SR II} selection criteria were plotted against the QCD multijets background as shown in Figure~\ref{fig:comp_SR1_SR2}. For both signal regions, the number of signal events was able to overcome the number of background events in the 700 - 800 GeV range. The {\fontfamily{cmtt}\selectfont SR II} result yields better sensitivity compared to {\fontfamily{cmtt}\selectfont SR I} for this sample point. This comparison demonstrates that the selection criteria in {\fontfamily{cmtt}\selectfont SR II} are more effective for heavier bino masses $(> \unit[750]{GeV})$.

As in Section~\ref{subsec:sri}, the selection criteria defined in {\fontfamily{cmtt}\selectfont SR II} were applied to the squark and bino mass ranges discussed in Section~3, and the resulting sensitivity is presented in the next subsection.

\subsubsection{Sensitivity Region for Signal Region II}

\begin{figure}[htbp]
  \centering
  \includegraphics[width = 0.8\linewidth]{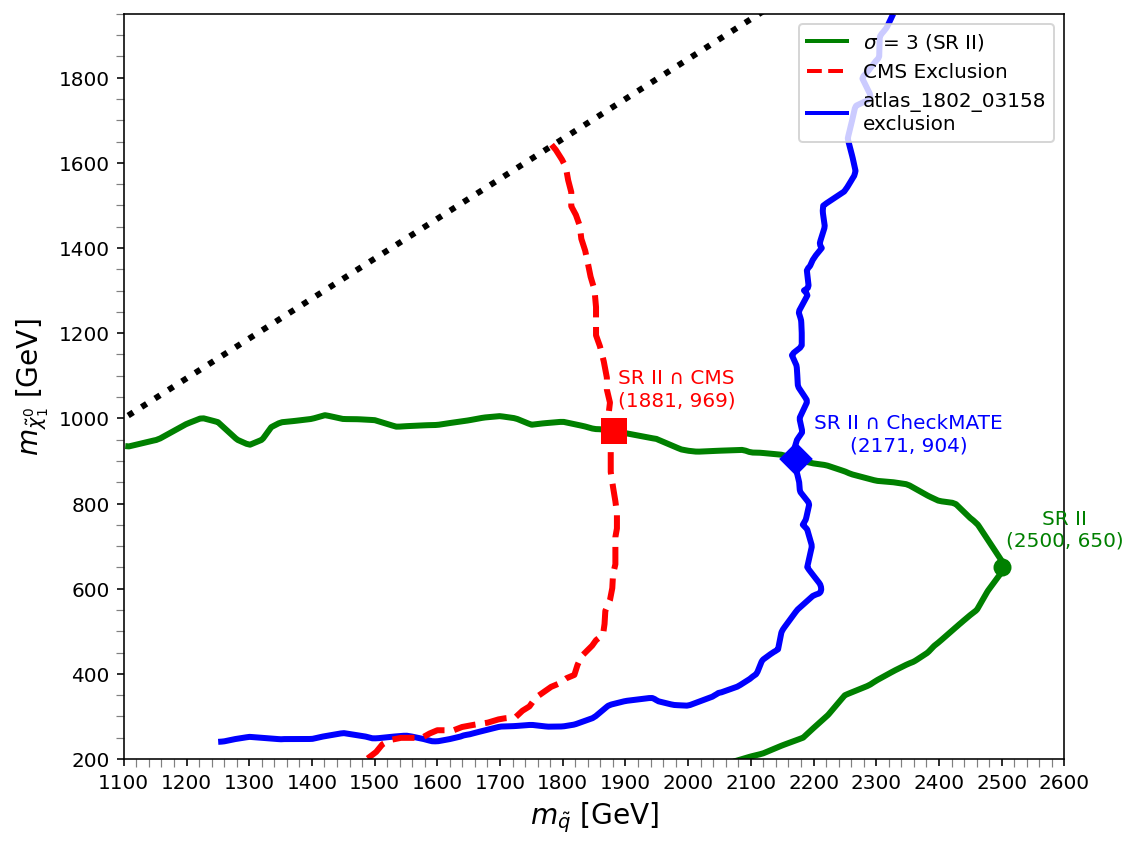}
  \caption{ Comparison of the {\fontfamily{cmtt}\selectfont SR II} sensitivity vs CMS exclusion. The events are normalized using an integrated luminosity of $\unit[138]{fb^{-1}}$.}
  \label{fig:CMS_SR2}
\end{figure}

The solid green line shown in Figure~\ref{fig:CMS_SR2} represents the sensitivity reach in the squark–bino mass phase space, obtained by applying the {\fontfamily{cmtt}\selectfont SR II} selection criteria.  This extension in sensitivity is compared with the dashed red line, representing the exclusion from the latest CMS stealth SUSY search of the same decay chain \cite{cms2023search}, and with the solid blue line, representing the {\fontfamily{cmtt}\selectfont atlas\_1802\_03158} analysis exclusion obtained from $\sqrt{s}=\unit[13]{TeV}$ searches using \textsc{CheckMATE}. With {\fontfamily{cmtt}\selectfont SR II}, the sensitivity extends the CMS exclusion to bino masses in the range $\unit[750]{GeV} < m_{\tilde{\chi}^0_1} \leq \unit[970]{GeV}$ and the {\fontfamily{cmtt}\selectfont atlas\_1802\_03158} exclusion to $\unit[750]{GeV} < m_{\tilde{\chi}^0_1} < \unit[900]{GeV}$, with sensitivity reaching squark masses of up to about $\unit[2500]{GeV}$ at $m_{\tilde{\chi}^0_1} = \unit[650]{GeV}$.

\begin{figure}[htbp]
  \centering
  \includegraphics[width = 0.9\linewidth]{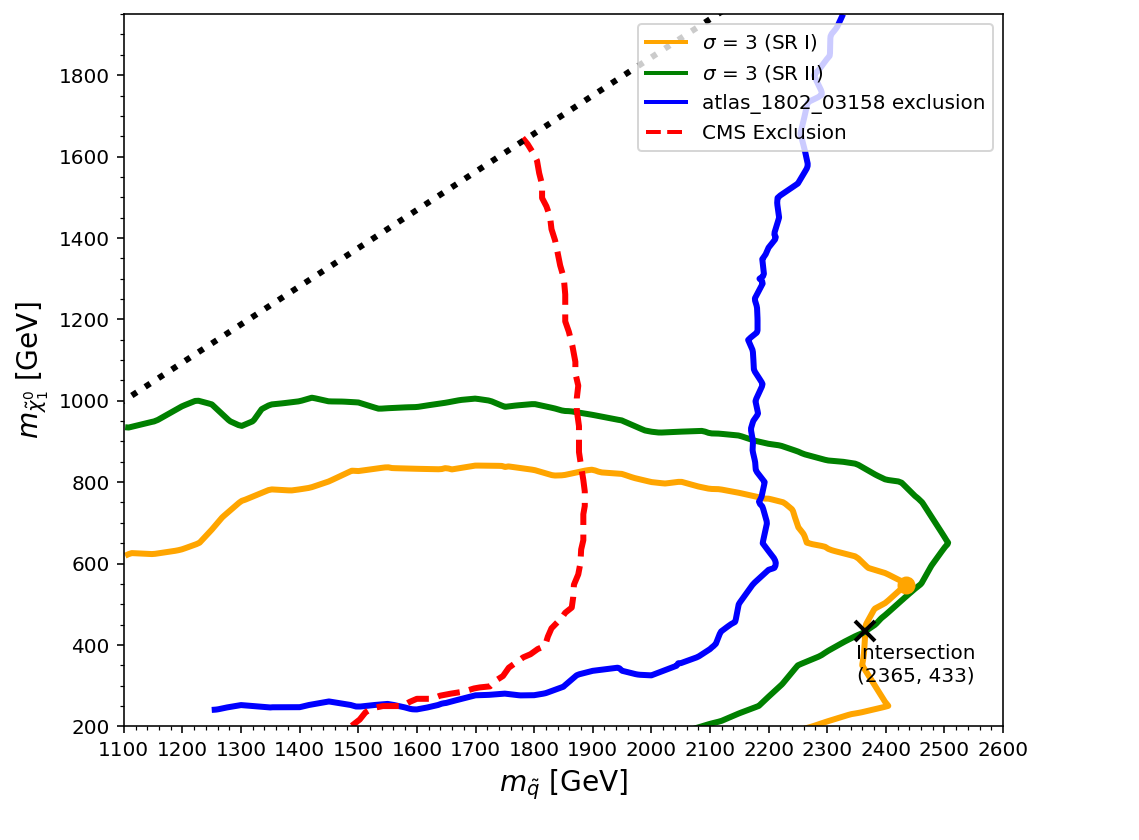}
  \caption{Combined sensitivity for {\fontfamily{cmtt}\selectfont SR I} and {\fontfamily{cmtt}\selectfont SR II}. The events are normalized using an integrated luminosity of $\unit[138]{fb^{-1}}$. }
  \label{fig:super_SR1_SR2}
\end{figure}

However, for bino masses beyond $\unit[900]{GeV}$, the selection criteria of {\fontfamily{cmtt}\selectfont SR II} are less sensitive. Above the point $(m_{\tilde{q}}, m_{\tilde{\chi}^0_1}) = (\unit[1881]{GeV}, \unit[969]{GeV})$, the CMS result provides a stronger exclusion in the squark-bino mass phase space. Also, the {\fontfamily{cmtt}\selectfont atlas\_1802\_03158} analysis yields a stronger exclusion above the $(m_{\tilde{q}}, m_{\tilde{\chi}^0_1}) = (\unit[2171]{GeV}, \unit[904]{GeV})$ point. Figure~\ref{fig:super_SR1_SR2} shows the combined sensitivity contours obtained from {\fontfamily{cmtt}\selectfont SR I} and {\fontfamily{cmtt}\selectfont SR II}. As shown in the figure, {\fontfamily{cmtt}\selectfont SR I} provides stronger sensitivity to squark masses at lighter bino masses. Beyond the point $(m_{\tilde{q}}, m_{\tilde{\chi}^0_1}) = (\unit[2365]{GeV}, \unit[433]{GeV})$, {\fontfamily{cmtt}\selectfont SR II} becomes more effective, offering improved sensitivity to squark masses for bino masses up to $\unit[900]{GeV}$.

The signal regions {\fontfamily{cmtt}\selectfont SR I} and {\fontfamily{cmtt}\selectfont SR II} demonstrated sensitivity up to $\unit[2500]{GeV}$ for bino masses below $\unit[900]{GeV}$. Beyond this point, the exclusion achieved by the ATLAS analysis remains more effective. Because this work adopts the initial conditions of a previous study~\cite{flores2020constraining}—specifically jet trimming and the requirement of large-radius jets with $p_T > \unit[450]{GeV}$—the selection criteria defined here are not sufficient to explore the region of parameter space with bino masses above $\unit[900]{GeV}$.

\begin{figure}[htbp]
  \centering
  \includegraphics[width = 0.8\linewidth]{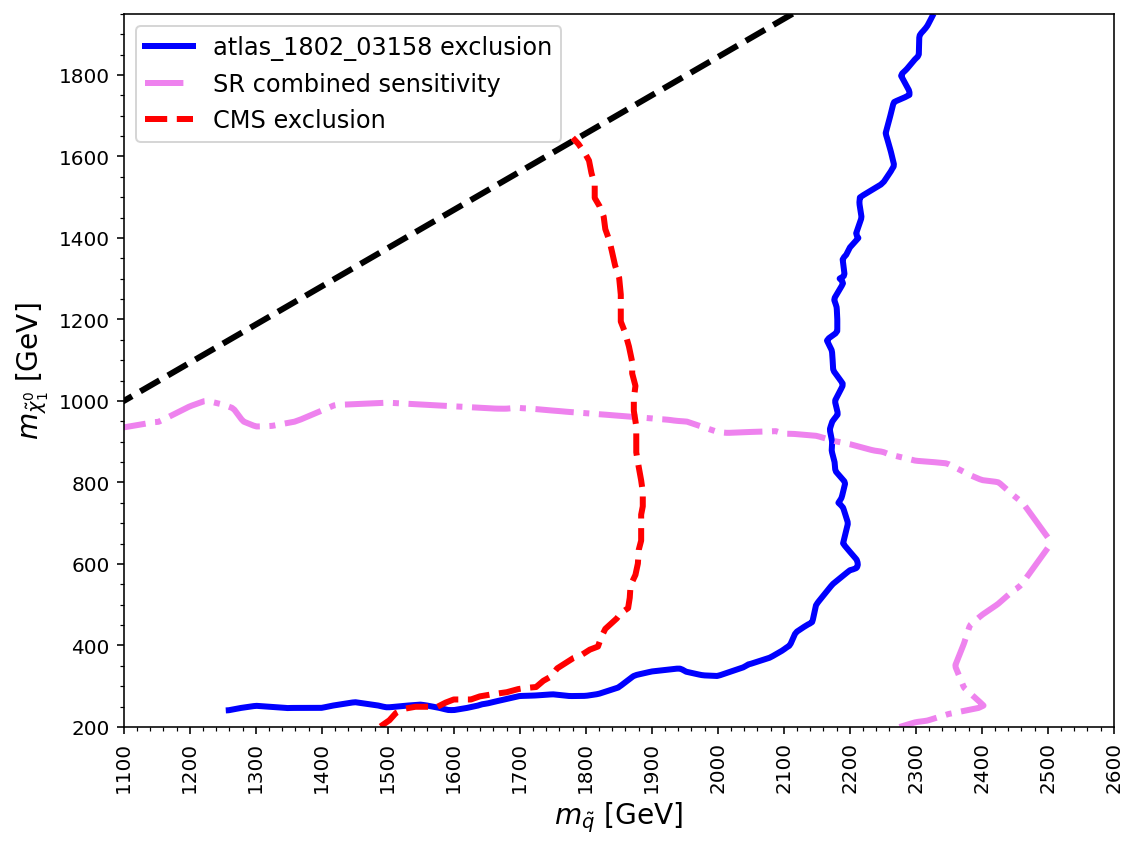}
  \caption{Combined exclusion plots of {\fontfamily{cmtt}\selectfont SR I} and {\fontfamily{cmtt}\selectfont SR II} together with the recasted ATLAS and CMS exclusion}
  \label{fig:combined_vs_CMS}
\end{figure}

Finally, Figure~\ref{fig:combined_vs_CMS} shows the combined sensitivity reach generated from the selection criteria in {\fontfamily{cmtt}\selectfont SR I} and {\fontfamily{cmtt}\selectfont SR II} (violet) vs the CMS exclusion of its latest stealth SUSY search (red) and the  {\fontfamily{cmtt}\selectfont atlas\_1802\_03158} analysis exclusion (blue). Although our analysis loses sensitivity at higher bino masses, it nicely complements the current strongest ATLAS and CMS exclusions, both taking care of the higher bino masses, while losing sensitivity in the lower regime. Hence, this analysis could potentially unveil or exclude stealth SUSY in the low missing energy terrain.

\section{Conclusion}
\label{sec:conclusion}

In this study, an improved sensitivity for squark-pair production was presented in the stealth SUSY model and compared with the latest CMS collaboration result which also investigated a similar decay chain. Also, it was shown that the {\fontfamily{cmtt}\selectfont atlas\_1802\_03158} analysis at $\unit[36]{fb^{-1}}$ integrated luminosity excluded a better range of squark masses compared to the CMS paper result. The results in this work suggest that for squark masses beyond $(>\unit[1850]{GeV})$, a signal should have been observed after Run 2 of the LHC with an integrated luminosity of $\unit[138]{fb}^{-1}$. The results showed that it is possible to extend the phase space of the squark-bino mass by utilizing kinematic cuts related to the properties of the decay products and the quantities measured in the detector ($\Delta R_{ij}$, multiplicity, etc.). 

For different bino mass ranges ($m_{\tilde{q}}$), signal regions were defined that correspond to a variety of kinematic variable cuts.The cuts were also tested against a dominant QCD multijets background. \textbf{Signal Region I} $(\unit[150]{GeV} \leq m_{\tilde{q}} < \unit[750]{GeV})$ used the following selection criteria: the multiplicity of leading-mass large-radius jet ($R = 1.0$) must be greater than 3, and the angular separation between the leading and next-leading photon ($\Delta R_{\gamma_1, \gamma_2}$) must be greater than 0.10. \textbf{Signal Region II} $(\unit[750]{GeV} \leq m_{\tilde{q}} < \unit[900]{GeV})$ defined the following selection criteria: the multiplicity of leading-mass large-radius jet ($R = 1.0$) must be greater than 2, the angular separation between the leading and next-leading photon ($\Delta R_{\gamma_1, \gamma_2}$) must be greater than 0.10, and the angular separation between the large-radius jet and the leading photon ($\Delta R_{j_1, \gamma_1}$) must be greater than 0.20.

Signal Region I was able to extend the sensitivity of squark masses up to $\unit[2400]{GeV}$ for a bino mass of around $\unit[550]{GeV}$ while Signal Region II was able to exclude up to $\unit[2500]{GeV}$ for a bino mass of $\unit[650]{GeV}$. Both signal regions demonstrate improved sensitivity to squark masses compared to the latest CMS search on stealth SUSY, as well as the most sensitive ATLAS exclusion, for bino masses up to approximately $\unit[900]{GeV}$.

\section*{Acknowledgements}
The authors thank the administration of Philippine Science High School- Main Campus for allowing them to use the school's computer laboratory to perform event simulation and analysis and the DOST- Human Resource Development Program (DOST-HRDP) for the graduate scholarship.


\clearpage
\onecolumn
\appendix

\section{List of \textsc{CheckMATE} Analyses}
\label{CheckMATE analysis}

\noindent
Table A.1 shows the list of $\sqrt{s}=\unit[13]{TeV}$ ATLAS and CMS analyses implemented in \textsc{CheckMATE} at the completion of this study.

\begin{center}
\centering
\begin{tabular}{|c|c|c|}
\hline
\multicolumn{1}{|c|}{\textbf{Analysis}}   & 
\multicolumn{1}{|c|}{\textbf{Description}} & 
\multicolumn{1}{c|}{$\mathcal{L}[\text{fb}^{-1}]$} \\ \hline
{\fontfamily{cmtt}\selectfont atlas\_1604\_01306} \cite{atlas2016photonmet}                 & photon + MET search at 13 TeV & 3.2\\ \hline
{\fontfamily{cmtt}\selectfont atlas\_1605\_09318} \cite{atlas2016gluinospair}               & b-jets $\geq 3$ + 0-1 lepton + ETMiss & 3.3  \\ \hline
{\fontfamily{cmtt}\selectfont atlas\_1609\_01599} \cite{atlas2016_ttZ_W}                    & ttV cross section measurement at 13 TeV & 3.2  \\ \hline
{\fontfamily{cmtt}\selectfont atlas\_1704\_03848} \cite{atlas2017_photonmet_dm}             & monophoton dark matter search & 36.1   \\ \hline
{\fontfamily{cmtt}\selectfont atlas\_1706\_03731} \cite{atlas2017_sssleptons}               & same-sign or 3 leptons RPC and RPV SUSY & 36.1    \\ \hline
{\fontfamily{cmtt}\selectfont atlas\_1708\_07875} \cite{atlas2018_chargino_neutralino_tau}  & electroweakino search with taus and MET & 36.1    \\ \hline
{\fontfamily{cmtt}\selectfont atlas\_1709\_04183} \cite{atlas2017_stop_allhadronic}         & stop pair production, 0 leptons & 36.1   \\ \hline
{\fontfamily{cmtt}\selectfont atlas\_1712\_02332} \cite{atlas2018_squarksgluinos}           & squarks and gluinos, 0 lepton, 2-6 jets & 36.1   \\ \hline
{\fontfamily{cmtt}\selectfont atlas\_1712\_08119} \cite{atlas2018_compressed_sleptons}      & electroweakinos search with soft leptons & 36.1    \\ \hline 
{\fontfamily{cmtt}\selectfont atlas\_1802\_03158} \cite{atlas2018_gmsb_photonic}            & search for GMSB with photons & 36.1  \\ \hline
{\fontfamily{cmtt}\selectfont atlas\_1803\_02762} \cite{atlas2018_two_three_leptons}        & electroweakino production in final states with two or three leptons & 36.1   \\ \hline
{\fontfamily{cmtt}\selectfont atlas\_1807\_07447} \cite{atlas2019_general_search}       & general search for new phenomena & 3.2   \\ \hline
{\fontfamily{cmtt}\selectfont atlas\_1908\_03122} \cite{atlas2019_bottom_squark_higgs_bjets}  & 0 leptons, 3 or more b-jets, sbottoms & 139  \\ \hline
{\fontfamily{cmtt}\selectfont atlas\_1908\_08215} \cite{atlas2020_charginos_sleptons}         & charginos/sleptons, 2 leptons + MET & 139   \\ \hline
{\fontfamily{cmtt}\selectfont atlas\_1909\_08457} \cite{atlas2020_same_sign_sleptons_gluinos}  & search for squarks and gluinos with same-sign leptons & 139  \\ \hline
{\fontfamily{cmtt}\selectfont atlas\_1911\_06660} \cite{atlas2020_stau_hadronic}                & search for direct stau production & 139       \\ \hline
{\fontfamily{cmtt}\selectfont atlas\_1911\_12606}\cite{atlas2020_compressed_spectra}             & search for sleptons and electroweakinos with soft leptons & 139     \\ \hline
{\fontfamily{cmtt}\selectfont atlas\_2004\_10894} \cite{atlas2020_direct_electroweakinos_photons} & EWino search in Higgs (diphoton) and MET& 139     \\ \hline
{\fontfamily{cmtt}\selectfont atlas\_2004\_14060} \cite{atlas2020_top_squark_allhadronic}         & stops, leptoquarks, 0 lepton & 139 \\ \hline
{\fontfamily{cmtt}\selectfont atlas\_2006\_05880} \cite{atlas2020_stop_higgs_Z}             & search for top squarks in events with a Higgs or a Z boson & 139         \\ \hline
{\fontfamily{cmtt}\selectfont atlas\_2010\_14293} \cite{atlas2021_squarksgluinos_139fb}     & search for squarks and gluinos in MET$\_$jet final states & 139        \\ \hline
{\fontfamily{cmtt}\selectfont atlas\_2101\_01629} \cite{atlas2021_squarksgluinos_lepton}    & squarks/gluinos, 1 lepton, jets, MET & 139        \\ \hline
{\fontfamily{cmtt}\selectfont atlas\_2102\_10874} \cite{atlas2021_monojet}                  & monojet search & 139        \\ \hline
{\fontfamily{cmtt}\selectfont atlas\_2103\_11684} \cite{atlas2021_four_leptons}             & search for SUSY in events with four or more leptons (gravitino SR) & 139         \\ \hline
{\fontfamily{cmtt}\selectfont atlas\_2106\_01676} \cite{atlas2021_chargino_neutralino_3lep} & electroweakinos, 3 leptons, WZ, Wh, on+off-shell & 139          \\ \hline
{\fontfamily{cmtt}\selectfont atlas\_2106\_09609} \cite{atlas2021_rpv_susy_leptons_jets}     & Search for RPV SUSY in final states with leptons and many jets & 139       \\ \hline
{\fontfamily{cmtt}\selectfont atlas\_2111\_08372} \cite{atlas2022_invisible_Higgs_Z}         & dark matter and invisible Higgs with Z boson & 139        \\ \hline
{\fontfamily{cmtt}\selectfont atlas\_2202\_07953} \cite{atlas2022_invisible_Higgs_VBF}       & invisible Higgs decays in VBF & 139       \\ \hline
{\fontfamily{cmtt}\selectfont atlas\_2209\_13935} \cite{atlas2023_sleptons_charginos}        & search for charginos and sleptons in 2-lepton final states & 139        \\ \hline
{\fontfamily{cmtt}\selectfont atlas\_2211\_08028} \cite{atlas2023_susy_missing_pt_bjets}     & search for gluinos decaying via 3rd gen; multi b-jets and MET & 139        \\ \hline
{\fontfamily{cmtt}\selectfont atlas\_conf\_2015\_082} \cite{atlas2015_conf082}     & leptonic Z + jets + ETMiss & 3.2               \\ \hline
{\fontfamily{cmtt}\selectfont atlas\_conf\_2016\_013} \cite{atlas2016_conf013}     & 4 top quark (1 lepton + jets, vector like quark search) & 3.2              \\ \hline
{\fontfamily{cmtt}\selectfont atlas\_conf\_2016\_050} \cite{atlas2016_conf050}     & 1-lepton + jets + ETmiss (stop) & 13.3              \\ \hline
{\fontfamily{cmtt}\selectfont atlas\_conf\_2016\_054} \cite{atlas2016_conf054}     & 1-lepton + jets + ETMiss (squarks and gluino) & 14.8             \\ \hline
{\fontfamily{cmtt}\selectfont atlas\_conf\_2016\_066} \cite{atlas2016_conf066}  & search for photons + jets + MET & 13.3              \\ \hline
{\fontfamily{cmtt}\selectfont atlas\_conf\_2016\_076} \cite{atlas2016_conf076}     & 2 leptons + jets + ETMiss & 13.3              \\ \hline
{\fontfamily{cmtt}\selectfont atlas\_conf\_2016\_096} \cite{atlas2016_conf096}     & 2-3 leptons + ETMiss (electroweakino) & 13.3               \\ \hline
{\fontfamily{cmtt}\selectfont atlas\_conf\_2017\_060} \cite{atlas2017_conf060}     & monojet search & 36.1              \\ \hline
{\fontfamily{cmtt}\selectfont atlas\_conf\_2019\_020} \cite{atlas2019_conf020}     & search for chargino-neutralino production with mass splittings near the electroweak scale & 139  \\ \hline
{\fontfamily{cmtt}\selectfont cms\_pas\_sus\_15\_011} \cite{cms2016_sfl_leptons_jets_met} & 2 leptons + jets + MET & 2.2  \\ \hline
{\fontfamily{cmtt}\selectfont cms\_sus\_16\_039} \cite{cms2018_charginos_neutralinos_multilep} & electrowekinos in multilepton final state & 35.9 \\ \hline
{\fontfamily{cmtt}\selectfont cms\_sus\_16\_025} \cite{cms2016_pas_sus16025}   & electroweakino and stop compressed spectra & 12.9  \\ \hline
{\fontfamily{cmtt}\selectfont cms\_sus\_16\_048} \cite{cms2018search}         & two soft opposite sign leptons & 35.9                 \\ \hline
{\fontfamily{cmtt}\selectfont cms\_sus\_19\_005} \cite{cms2019_pas_sus19005} & hadronic final states with $M_{T2}$ & 137   \\ \hline
{\fontfamily{cmtt}\selectfont cms\_1908\_04722} \cite{cms2019_susy_jets_met} & hadronic final states with HT, post-fit and simple fitting & 137    \\ \hline
{\fontfamily{cmtt}\selectfont cms\_2107\_13201} \cite{cms2021_jets_met}    & monojet with multibin & 137      \\ \hline
{\fontfamily{cmtt}\selectfont cms\_2205\_09597} \cite{tumasyan2023search}  & search for electroweakinos in hadronic final states & 137    \\ \hline
\end{tabular}
\end{center}

\clearpage
\twocolumn


\bibliographystyle{elsarticle-num} 
\bibliography{example}






\end{document}